\documentclass[preprint]{aastex}

\shorttitle{The density and velocity structures of B335 core}
\shortauthors{Kurono et al.}

\begin{document}

\title{
UNVEILING THE DETAILED DENSITY AND VELOCITY STRUCTURES OF THE PROTOSTELLAR CORE B335
}

\author{Yasutaka Kurono\altaffilmark{1}, 
    Masao Saito\altaffilmark{1,2},
    Takeshi Kamazaki\altaffilmark{1,2},
    Koh-Ichiro Morita\altaffilmark{1,2},
and
    Ryohei Kawabe\altaffilmark{1,2}}
\altaffiltext{1}{The Chile Observatory, National Astronomical Observatory
of Japan, Osawa 2-21-1, Mitaka, Tokyo 181-8588, Japan}
\altaffiltext{2}{Joint ALMA Observatory, Alonso de Cordova 3107 Vitacura, Santiago 763 0355, Chile}
\email{yasutaka.kurono@nao.ac.jp}

\begin{abstract}
We present an observational study of the protostellar core B335 harboring 
a low-mass Class 0 source.
The observations of the ${\rm H^{13}CO^+}(J=1$--$0)$ line emission were 
carried out using the Nobeyama 45 m telescope and Nobeyama Millimeter 
Array.
Our combined image of the interferometer and single-dish data depicts 
detailed structures of the dense envelope within the core.
We found that the core has a radial density profile of 
$n(r)\propto r^{-p}$ and a reliable difference in the power-law indices
between the outer and inner regions of the core: 
$p\approx 2$ for $r\gtrsim4000\,{\rm AU}$ and 
$p\approx 1.5$ for $r\lesssim 4000\,{\rm AU}$.
The dense core shows a slight overall velocity gradient of 
$\sim 1.0\,{\rm km\,s^{-1}}$ over the scale of $20{,}000\,{\rm AU}$ 
across the outflow axis.
We believe that this velocity gradient represents a solid-body-like
rotation of the core.
The dense envelope has a quite symmetrical velocity structure 
with a remarkable line broadening toward the core center,
which is especially prominent in the position--velocity diagram across 
the outflow axis. 
The model calculations of position--velocity diagrams do a good job of 
reproducing observational results using the collapse model of an 
isothermal sphere 
in which the core has an inner free-fall region and an outer region 
conserving the conditions at the formation stage of a central stellar 
object.
We derived a central stellar mass of $\sim 0.1M_\sun$, and suggest
a small inward velocity,
$v_{r \geqslant r_{\rm inf}}\sim 0\,{\rm km\,s^{-1}}$ in the outer core at
$\gtrsim 4000\,{\rm AU}$.
We concluded that our data can be well explained by gravitational 
collapse with a quasi-static initial condition, such as Shu's model, 
or by the isothermal collapse of a marginally critical Bonnor--Ebert 
sphere.
\end{abstract}

	
\section{INTRODUCTION}
\label{sec:1}
In order to understand the formation processes of low-mass stars, it is 
important to investigate the properties of dense 
($\sim 10^5\,{\rm cm^{-3}}$) cores in molecular clouds.  
Such compact ($\sim 0.1\,{\rm pc}$) cores supply material to 
newly forming stars through dynamical gravitational collapse, however,
detailed physical processes are still uncertain.
One of the investigative approaches is to derive the detailed 
density and velocity structures from observations of (pre-)protostellar 
cores which are expected to retain more information
than Class I/II objects for the initial conditions of gravitational 
collapse \citep{1993ApJ...406..122A,1999ApJ...518..334S,2006ApJ...653.1369F}. 

Dust continuum emission imaging at millimeter and submillimeter 
wavelengths using single-dish radio telescopes has revealed the radial 
density profiles, $\rho(r)$, of (pre-)protostellar cores 
\citep{1994MNRAS.268..276W,1999MNRAS.305..143W,1996A&A...314..625A,2000ApJS..131..249S}.
Recent investigations have demonstrated that the profiles of Class 0/I 
sources can be fitted by single power-law profiles over a wide range of 
radii \citep[e.g.,][]{2000ApJS..131..249S}.
\citet{2002ApJ...575..337S} modeled Class 0 source maps using a single 
power-law density distribution $\rho(r)\propto r^{-p}$ and found that 
most of them can be well fitted with a power-law 
index of $p\sim 1.8$.
Gas kinematics in dense cores have been investigated by
molecular line observations.
It has been shown that asymmetric double-peaked profiles of 
optically thick lines detected toward star-forming cores are considered to 
be a signature of collapse motion.
Extensive surveys of such blue-skewed spectra in starless cores have been
carried out by 
\citet{1999ApJ...526..788L,2001ApJS..136..703L,2004ApJS..153..523L}.
Furthermore, \citet{1998ApJ...504..900T} conducted the profile fitting 
using a simple two-layer radiative transfer model and suggested an inward 
motion of subsonic speed ($0.02$--$0.1\,{\rm km\,s^{-1}}$) that extended 
to ${\rm 0.1\,pc}$ in the pre-protostellar core L1544.
The velocity structures of dense cores, including not only infalling
motion but also rotation, have also been investigated 
using first and second moment images of line emission
taken with single-dish telescopes and interferometers 
\citep[e.g.,][]{2011ApJ...740...45T}.
Statistical studies for the rotation of dense cores were conducted 
using ${\rm NH_3}$ and ${\rm N_2H^+}$ line emissions by 
\citet{1993ApJ...406..528G} and \citet{2002ApJ...572..238C}, in which the 
typical velocity gradients are found to be 
$1$--$2\,{\rm km\,s^{-1}\,pc^{-1}}$.
On the other hand, \citet{2007ApJ...669.1058C} observed 
${\rm N_2H^+}(J=1$--$0)$ line emission toward nine low-mass protostellar 
envelopes down to $\sim 1000 {\rm AU}$ scales.
The mean velocity gradient estimated in their samples is 
$\sim7\,{\rm km\,s^{-1}\,pc^{-1}}$, which is much larger than the velocity
gradients of dense cores.
\citet{2011ApJ...740...45T} analyzed the kinematics of 17 protostellar 
systems. 
They found that the velocity gradients obtained with interferometric 
data ($\sim 8.6\,{\rm km\,s^{-1}\,pc^{-1}}$) are considerably larger than 
those that also have single-dish data 
($\sim 2.3\,{\rm km\,s^{-1}\,pc^{-1}}$), which indicates accelerating 
infall and spinning-up rotational velocities toward the core center.

The environments surrounding forming stars are composed of
structures with different scales: circumstellar disks
($\lesssim 100\,{\rm AU}$), infalling envelopes ($\sim 1000\,{\rm AU}$), 
and dense cores ($\sim 10{,}000\,{\rm AU}$).
Recently, an approach that combines data obtained with single-dish
telescopes and interferometers has been widely used to investigate the 
physics in protostellar systems
\citep[e.g.,][]{2006ApJ...653.1369F,2007ApJ...662..431T,2011ApJ...742...57Y}.
\citet{2011ApJ...742...57Y} performed ${\rm C^{18}O} (J=2$--$1)$ and 
${\rm CS} (J=7$--$6)$ observations toward the protostellar
envelope of B335 with the Submillimeter Array and single-dish telescopes, 
and imaged by combining those data.
They derived the specific angular momentum of the envelope and found that 
specific angular momenta tend to be larger as evolution progresses by 
comparing with other Class 0, I, and II sources. 

Theoretically, two extreme models for the core evolution have been proposed
for isolated low-mass star formation.
The similarity solution of Larson--Penston describes the density evolution
of isothermal gas spheres \citep{1969MNRAS.145..271L,1969MNRAS.144..425P}.
When a central object is formed ($t=0$), the gas sphere
reaches the density profile of $\rho(r) = 4.4(c_{\rm s}^2/2\pi G)r^{-2}$ 
and velocity field of $v(r)=3.3\,c_{\rm s}$,
where $c_{\rm s}$ is the isothermal sound speed and $G$ is the 
gravitational constant.
This model of the Larson--Penston solution is referred to as ``runaway''
collapse.
On the other hand, the isothermal similarity solution proposed by
\citet{1977ApJ...214..488S} describes a core that is slowly increasing 
its central density through ambipolar diffusion while maintaining
kinematic balance (i.e., $v(r)=0$), and moving toward dynamical collapse. 
The density profile achieves a singular isothermal sphere, 
$\rho(r) = (c_{\rm s}^2/2\pi G)r^{-2}$, at $t=0$, 
which is the initial condition of dynamical collapse after 
protostar formation.
This model is the most static and is referred to as ``inside-out'' 
collapse.
Moreover, extensions for $t>0$ of the Larson--Penston solution and 
generalization were developed by \citet{1977ApJ...218..834H} and 
\citet{1985MNRAS.214....1W}.
At later times ($t>0$), in both of the solutions, the density and
velocity structures attain a free-fall profile, $\rho\propto r^{-3/2}$ and
$v\propto r^{-1/2}$, respectively, from the center to the outside, with a
sound speed for the Shu solution that accompanies the rarefaction wave and
with a supersonic velocity for the Larson--Penston solution.
The mass infall rate is predicted to be $\sim c_{\rm s}^3/G$ for the Shu
solution and 48 times higher than this for the Larson--Penston solution.

In this paper, we present an observational study of the dense core 
associated with a Class 0 protostar within B335.
The Bok Globule B335, otherwise known in the
literature as CB199 in the catalog of \citet{1988ApJS...68..257C} or L663,
is a typical low-mass star-forming region.
B335 appears as an opaque dark cloud on optical images and is one of the 
best candidates for studying the initial conditions of star formation,
because it is isolated from other star-forming regions and is near
the Sun.
In this paper, we adopt a distance of $150\,{\rm pc}$
\citep{2008ApJ...687..389S}.
B335 contains a far-infrared (FIR) source, IRAS 19347+0727, which is 
bright in submillimeter wavelengths \citep{1990MNRAS.243..330C} and 
shows a combination of characteristics that indicates one of the clearest 
examples of very young stars.
This FIR source is associated with the dense molecular gas envelope 
\citep{1987ApJ...313..320F,1989A&A...223..258M,1991ApJ...374..177H,1999ApJ...518..334S,2010ApJ...710.1786Y,2011ApJ...742...57Y}, and is classified as a Class 0 protostellar object based on 
its spectral energy distribution \citep{1994ASPC...65..197B}. 
This source is considered to be the driving source of the bipolar outflow 
\citep{1982ApJ...256..523F,1988ApJ...327L..69H,1992ApJ...390L..85H,1993ApJ...414L..29C,2010ApJ...710.1786Y}, extending from east to west 
with an inclination angle from the sky plane of $\sim 10^\circ$ and
an opening angle of $\sim 45^\circ$.
Toward the IRAS source in B335, 
\citet{1993ApJ...404..232Z,1994ApJ...421..854Z} have obtained
blue-skewed line profiles in the ${\rm CS}(J=2$--$1,~3$--$2,~{\rm 
and}~5$--$4)$ and ${\rm H_2CO}(J=2_{12}$--$1_{11} ~{\rm and} ~3_{12}$--$2_{11})$
line emissions, 
which can be explained by a spherically symmetric inside-out collapse 
model \citep{1977ApJ...214..488S}.
Moreover, \citet{1995ApJ...448..742C} computed radiative transfer 
using a Monte Carlo method with a model of inside-out collapse.
They confirmed that the double-peaked line profiles indicate an infall 
motion and derived the physical parameters of the collapsing envelope.
\citet{1995ApJ...451L..75V} made interferometric observations in the ${\rm 
CCS}$ line and also interpreted the kinematics as an infalling motion,
whereas the interferometric images of ${\rm CS} (J=5$--$4)$ by 
\citet{2000ApJ...544L..69W} were found to be dominated by clumps 
associated with the outflow cavity.

As mentioned above, the detailed kinematics of infall motion provide
crucial information about the properties of the gravitational collapse of 
the core, which we can compare with the theoretical pictures and 
distinguish between them.
Few observational studies until now, however, have been capable of 
unveiling the velocity structures within dense cores in the early phases 
of star formation over a wide range of spatial scales.
We carried out the observations with the Nobeyama 45 m telescope and 
the Nobeyama Millimeter Array (NMA) in the ${\rm H^{13}CO^+}(J=1$--$0)$
line emission.
Observations with the 45 m telescope trace the overall density and 
kinematic structures of the star-forming core.
For the high-resolution observations with NMA, we combined the data 
with the 45 m telescope data to image the structures of the dense core 
over a wider spatial frequency range down to the inner dense envelope 
scales.

\section{OBSERVATIONS AND DATA ANALYSIS}
\label{sec:2}

\subsection{Nobeyama 45 m Telescope}
\label{subsec:2.1}

The mapping observations of the ${\rm H^{13}CO^+}(J=1$--$0)$ line 
emission at $86.75433\,{\rm GHz}$ toward the B335 region were carried 
out during 2006 March--May.
We used the 25 Beam Array Receiver System 
\citep[BEARS;][]{2000SPIE.4015..237S,2000SPIE.4015..614Y},
which consists of superconductor--insulator--superconductor 
(SIS) receivers on a $5\times 5$ grid with a separation of $41\farcs1$, 
in the double sideband (DSB) mode.
The observations were made in the On-The-Fly (OTF) mode of the 
Nobeyama Radio Observatory (NRO) 45 m telescope 
\citep{2008PASJ...60..445S}.
The mapping center was placed at the IRAS source 19347+0727 
associated with B335, ${\rm R.A.}=19^{\rm h} 34^{\rm m} 35\fs1$, 
${\rm decl.}=+07^\circ 27' 20\farcs0$ (B1950),
and the mapping region
was $\approx 4\times 4\,{\rm arcmin^2}$.
We obtained five OTF-scan data over this region in the R.A. 
and decl. directions in each and merged them to create a final map 
cube.
At $87\,{\rm GHz}$, the half-power beam width and main-beam efficiency 
were $18\farcs5$ and $0.5$, respectively. 
As the back end, we used 25 sets of $1024$ channel autocorrelators 
(ACs), with a frequency resolution of $31.25\,{\rm kHz}$ corresponding
to a velocity resolution of $0.108\,{\rm km\,s^{-1}}$.
The system noise temperatures were between $200\,{\rm K}$ and 
$280\,{\rm K}$.
We took the off position of ${\rm R.A.}=19^{\rm h}
36^{\rm m} 35\fs2$, ${\rm decl.}=+07^\circ 27' 22\farcs0$ (B1950)
and used the standard  chopper wheel method to convert the receiver output 
into $T_{\rm A}^\ast$ intensity scale.
The telescope pointing was checked once an hour by five-point observations
of the SiO maser from RT Aql in the $43\,{\rm GHz}$ using the SIS 
receiver (S40).
The pointing accuracies were within $3\arcsec$ for the ${\rm H^{13}CO^+}$ 
observations.

For the ${\rm H^{13}CO^+}$ data obtained with the OTF observations, 
IDL\footnote{
The Interactive Data Language
}-based reduction software, NOSTAR 
(Nobeyama OTF Software Tools for Analysis and Reduction),
was used for flagging, baseline subtraction, 
and making the map cube.
We corrected for relative gain differences among the 25 beams using 
correction factors provided by NRO, and then made the baseline fitting and 
subtraction.
We made a three-dimensional image cube that has an effective spatial 
resolution of $25\farcs1$ and an achieved 
noise level in channel images ($1\sigma$) of $95\,{\rm mK}$ in $T_A^\ast$ 
with a velocity resolution of $0.108\,{\rm km\,s^{-1}}$.

We also observed three inversion spectra of 
${\rm NH_3}(J,K)=(1,1)$, $(2,2)$, 
and $(3,3)$ at $23\,{\rm GHz}$ toward the IRAS source associated with B335
using the HEMT receiver (H22) during 2006 March--May.
At ${\rm 23\,GHz}$, the half-power beam width and main-beam efficiency
were $78\arcsec$ and 0.82, respectively.
We used eight sets of 2048 channel acousto-optical spectrometers
(AOSs) that have a frequency resolution of $37\,{\rm kHz}$, corresponding
to the velocity resolution of $0.48\,{\rm km\,s^{-1}}$.
The typical system noise temperatures were in a range of $120$--$280\,{\rm K}$
during the observations.
The achieved noise level in the spectrum was about $30\,{\rm mK}$ in 
$T_A^\ast$, with a velocity resolution of 
$0.48\,{\rm km\,s^{-1}}$.
The pointing accuracies were within $7\arcsec$ for the ${\rm NH_3}$ 
observations.

The observational parameters are summarized in Table \ref{tbl:1} 
for the 45 m telescope observations.

\subsection{Nobeyama Millimeter Array}
\label{subsec:2.2}
The aperture synthesis observations were carried out using the six-element
NMA. 
We observed the ${\rm H^{13}CO^+}(J=1$--$0)$ line in the period from 
2005 December to 2006 April. 
The phase tracking center for our observations was set at the position of
the IRAS source in B335.
The field of view (i.e., the primary beam size) of the $10\,{\rm m}$
dishes was about $79\arcsec$ at $87\,{\rm GHz}$.
Our observations were conducted with two array configurations (D and 
C configurations). 
For the maximum scale detectable in our observations, 
$\lambda/B_{\rm min} \approx 71\farcs1(\nu/87\,{\rm GHz})^{-1}(B_{\rm min}/10\,{\rm m})^{-1}$, 
the expected brightness recovered at the phase center is 
$\approx 2.8\%$ \citep{1994ApJ...427..898W} when a Gaussian brightness 
distribution having a FWHM of $\theta_{\rm FWHM} = \lambda/B_{\rm min}$ 
is assumed.
We used the SIS receivers as the front end and the digital 
spectro-correlator, New-FX, which has 1024 channels and bandwidth of 
$32\,{\rm MHz}$ (achieved 
velocity resolution was $0.108\,{\rm km\,s^{-1}}$), as the back end. 
We also obtained the continuum data simultaneously with the other
spectro-correlator, Ultra-Wide Band Correlator
\citep[UWBC;][]{2000PASJ...52..393O}, which has 128 
channels in the $1024\,{\rm MHz}$ bandwidth mode.
The system temperatures were typically in the range of 
$150$--$200\,{\rm K}$ in DSB.

We used 3C345 and 3C454.3 as passband calibrators and B1923+210 as a 
gain calibrator.
The absolute flux densities were calibrated using
the flux density of B1923+210 determined by the bootstrapping method 
with Uranus and Neptune. 
The measured flux density of B1923+210 was $1.26$--$1.61\,{\rm Jy}$ at 
$87\,{\rm GHz}$ during the observation period.
The uncertainty in our flux calibration was expected to be less than 10\%.

The NMA visibility data were processed (calibration, flagging, and 
continuum subtraction) using the UVPROCII package.
For the continuum data, we merged the visibilities in each sideband
obtained with the UWBC to construct an image.
The calibrated visibility data set was processed through imaging and 
image reconstruction with the MIRIAD package.
The synthesized beam size is $5\farcs5\times4\farcs3$ (position angle 
(P.A.) $=-32\fdg7$) at $87\,{\rm GHz}$ with natural weighting.
Imaging noise levels ($1\sigma$) of ${\rm H^{13}CO^+}$ and 
$87\,{\rm GHz}$ continuum data are $40\,{\rm mJy\,beam^{-1}}$, with a
velocity resolution of $0.108\,{\rm km\,s^{-1}}$ and 
$0.56\,{\rm mJy\,beam^{-1}}$, respectively.

The parameters for the NMA observations are summarized in 
Table \ref{tbl:2}.

\subsection{Combining the 45 m Telescope and the NMA Data}
\label{subsec:2.3}
We applied a data combining technique of interferometer and single-dish
data in the Fourier domain ($u$--$v$ domain) for our NMA and 45 m 
telescope ${\rm H^{13}CO^+}$ data.
We made the 45 m telescope image cube using a gridding kernel of 
Spheroidal \citep{2008PASJ...60..445S} with a grid size of $7\farcs92$,
which is the Nyquist spacing of a dish diameter of $45\,{\rm m}$ at 
$87\,{\rm GHz}$.
The velocity separation between the image channels is
$0.108\,{\rm km\,s^{-1}}$, which is the same as that 
of the NMA observations.
The short spacing information in the $u$--$v$ domain including the 
zero-baseline within the diameter of the primary antenna of NMA 
($=10\,{\rm m}$), which cannot inherently be obtained with the
NMA observations, can be complemented with the 45 m telescope data.

The fundamental theory and detailed algorithm of data combining are 
described in \citet{2009PASJ...61..873K}.
In this method, we generated pseudo-visibilities from the Fourier 
transformed single-dish image data, which are deconvolved by a Gaussian 
beam with an FWHM of $25\farcs1$ and multiplied by a primary beam of NMA
approximated by a Gaussian function with an FWHM of $79\arcsec$.
The single-dish visibilities that were generated were combined with the 
NMA data in the $u$--$v$ domain.
As a result, we can make synthesis images for a high spatial dynamic range
and a high spatial resolution while recovering missing large-scale fluxes. 
We applied the data optimizations to sensitivities and relative weights
between 45 m and NMA data \citep{2009PASJ...61..873K}.
We finally obtained a combined image of B335 in the ${\rm H^{13}CO^+}$ 
line emission with a synthesized beam size of 
$5\farcs6\times 4\farcs4$ (P.A. of $32\fdg7$) and with a total
flux of $32.4\,{\rm Jy\,km\,s^{-1}}$ that is comparable to that of the
45 m telescope image, $30.0\,{\rm Jy\,km\,s^{-1}}$.

Furthermore, Y. Kurono et al. (2013, in preparation) describe the schemes
that are applied to our observed data, and demonstrate the importance of 
relative (1) flux scaling, (2) sensitivity, and (3) weighting between 
interferometer and single-dish data in order to obtain reliable results.

\section{RESULTS}
\label{sec:3}

\subsection{${\rm NH_3}$ Line Emission}
\label{subsec:3.1}

\begin{figure}[htbp]
\includegraphics[width=.8\linewidth]{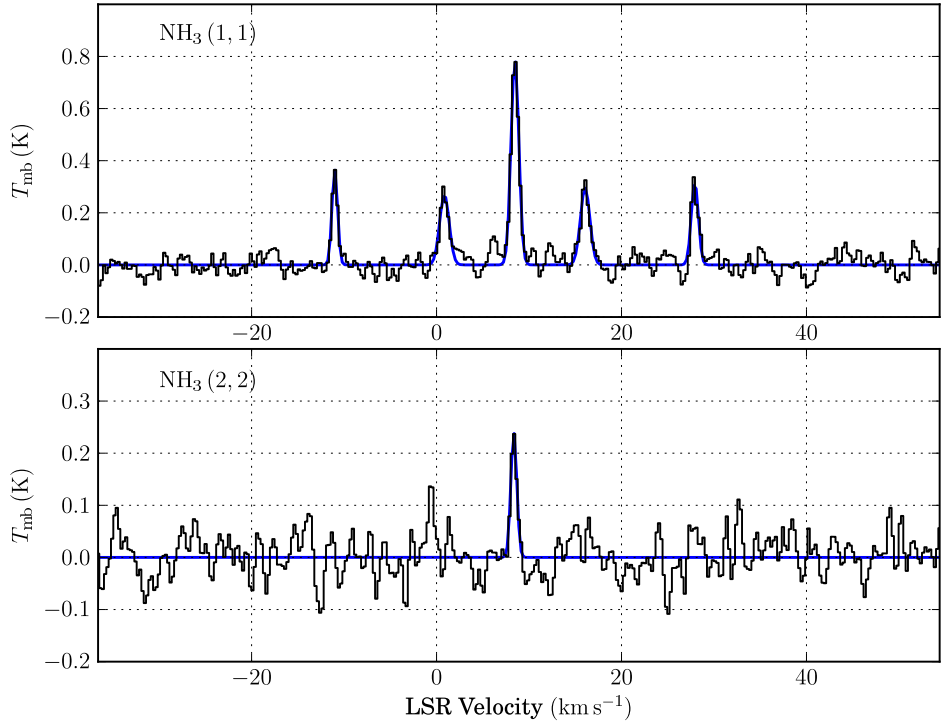}
\caption[
${\rm NH_3}$ line profiles observed with the Nobeyama 45 m 
telescope toward B335.
]
{
${\rm NH_3}(J,K) = (1,1)$ (upper panel) and $(J,K) = (2,2)$ (lower panel)
line profiles (black histogram) observed with the Nobeyama 45 m 
telescope toward the B335. 
Blue curves represent the best-fit results of Gaussian fitting
(Section \ref{subsec:3.1}).
The rms noise levels are $42\,{\rm mK}$ in $T_{\rm mb}$ for
both spectra.
}
\label{fig:1}
\end{figure}

The spectrum of the ${\rm NH_3}(J,K)=(1,1)$ transition, which consists of 
five hyperfine groups, and the main component of the $(2,2)$ transition 
were obtained with good signal-to-noise ratios for the 45 m telescope.
Figure \ref{fig:1} shows the spectra of the ${\rm NH_3}(J,K)=(1,1)$ and 
$(2,2)$ transitions obtained at the position of the IRAS source in B335. 
No emission in the $(3,3)$ transition was detected. 

To estimate the optical depth and gas kinetic temperature, we analyzed the
hyperfine structures that are caused by the electric quadrupole moment of 
the nitrogen nucleus.
We derived the peak main-beam temperatures ($T_{\rm mb}$) and intrinsic 
velocity widths ($\Delta v$) by fitting each line component with a 
Gaussian function. 
We assumed that all hyperfine components have equal beam filling factors
and excitation temperatures.
We estimated the optical depth of the ${\rm NH_3}(1,1)$ main component 
[$\tau(1,1,m)$], the rotational temperature [$T_{\rm rot}(2,2;1,1)$]
that describes the relative population between the $(1,1)$ and $(2,2)$ 
levels, and the kinetic temperature ($T_{\rm k}$) by following the 
analysis in a previous work by \citet{1992ApJ...388..467M}.
With the parameters of the best-fit results (summarized in Table \ref{tbl:3}) shown by the blue curves in 
Figure \ref{fig:1}, we obtained $\tau(1,1,m)=1.5 \pm 
0.9$, $T_{\rm rot}(2,2;1,1) = 12.9 \pm 1.1\,{\rm K}$, and $T_{\rm k} = 
15.1\pm 1.5\,{\rm K}$.
The rotational temperature is almost equal to the kinetic temperature for 
the range of $T_{\rm rot}(2,2;1,1)\lesssim 20\,{\rm K}$ 
\citep{1988MNRAS.235..229D}.
Therefore, we determined that the B335 core has a mean kinetic temperature 
of $\sim {\rm 15\,K}$ over the beam area of the 45 m telescope 
($\sim 80\arcsec$), and used this temperature to estimate the isothermal 
sound speed and the column density from the molecular line data.

\subsection{87GHz Continuum Emission}
\label{subsec:3.2}

\begin{figure}[htbp]
\includegraphics[width=.75\linewidth]{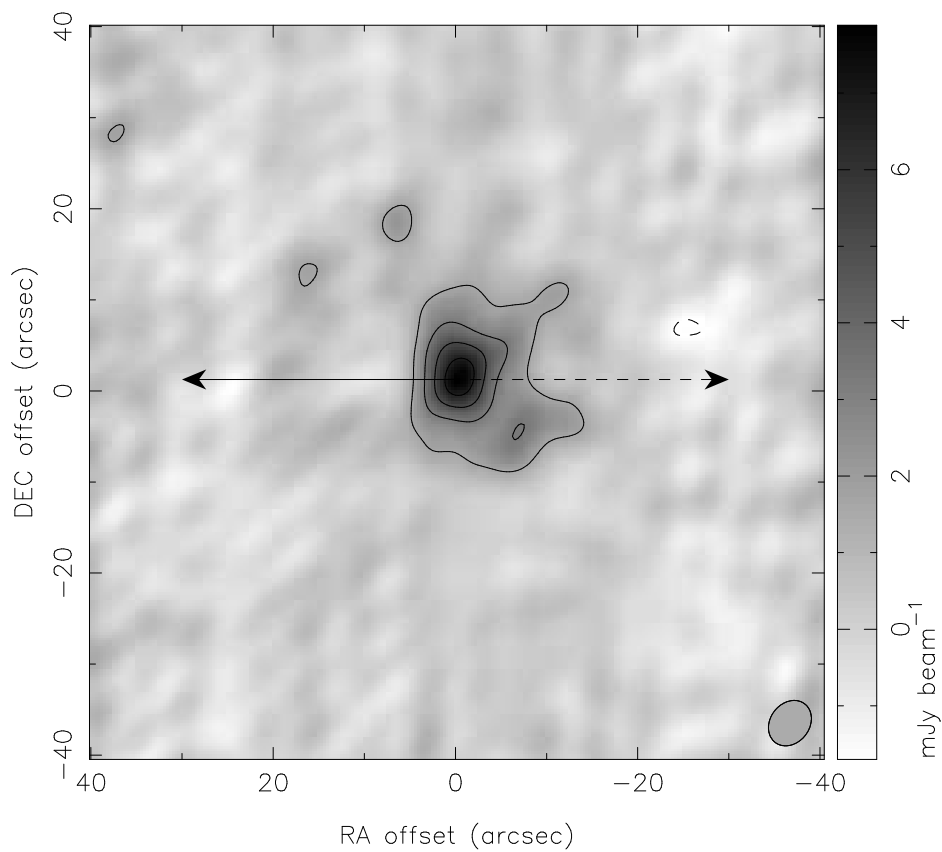}
\caption[
$87\,{\rm GHz}$ continuum images of B335.
]
{
$87\,{\rm GHz}$ continuum image toward the IRAS source in B335 
observed with the NMA.
Contours start from the $3\sigma$ level with a $3\sigma$ interval,
where $1\sigma=0.56\,{\rm mJy\,beam^{-1}}$.
Dashed contours show the $-3\sigma$ level.
Solid and dashed arrows in the image indicate the outflow directions 
of blue and red lobes, respectively.
The ellipse at the bottom right corner indicates the synthesized beam 
size ($5\farcs33\times 4\farcs31$ with a position angle of $-37\fdg2$).
}
\label{fig:2}
\end{figure}

We detected $87\,{\rm GHz}$ continuum emission with the NMA toward the 
IRAS source in B335, as shown in Figure \ref{fig:2}. 
The peak position was measured to be ${\rm R.A.}=19^{\rm h}
34^{\rm m} 35\fs2$, ${\rm decl.}=+07^\circ 27' 23\farcs8$ (B1950), 
which is consistent with the peak position of the $1.3\,{\rm mm}$ 
continuum images \citep{1999ApJ...526..833H, 2010ApJ...710.1786Y}.
The $87\,{\rm GHz}$ continuum image shows an elongated emission structure 
perpendicular to the outflow lying from east to west
\citep{1988ApJ...327L..69H}.
Furthermore, a cavity-like distribution in the red-lobe side of the 
molecular outflow can be seen which is quite similar to that in the 
$1.3\,{\rm mm}$ continuum image shown by \citet{2010ApJ...710.1786Y} and 
supports their suggestion that this distribution traces the wall of the 
outflow cavity.
The beam deconvolved size, peak intensity, and total flux density (above 
the $3\sigma$ contour) are $9\farcs2\times 6\farcs4$ 
(corresponding to $1380\, {\rm AU} \times 960\,{\rm AU}$), 
$7.89\pm 0.56\,{\rm mJy\,beam^{-1}}$, and
$33.1\pm 0.17\,{\rm mJy}$, respectively.

As mentioned in \citet{1993ApJ...414L..29C}, contribution from free--free
emission is considered to be negligible when taking into account the 
extrapolation from the flux at $3.6\,{\rm cm}$.
Thus, we determine that the $87\,{\rm GHz}$ continuum emission comes from 
a dust envelope surrounding a protostellar object.
Under the condition of being optically thin for thermal dust emission, the 
dust envelope mass ($M_{\rm env}$) was estimated using the equation 
$M_{\rm env} = S_\nu d^2 / \kappa_\nu B_\nu(T_{\rm d})$, 
where $S_\nu$ is the total flux density, $\kappa_\nu$ is the dust mass 
opacity coefficient, $T_{\rm d}$ is the 
dust temperature, $d$ is the source distance, and $B_\nu$ is the Planck 
function.
By combining our $87\,{\rm GHz}$ and image band $99\,{\rm GHz}$ 
measurements with the flux densities at millimeter wavelengths estimated 
by \citet{1983ApJ...274L..43K}, \citet{1990MNRAS.243..330C}, and 
\citet{1992ApJ...390L..85H}, 
we obtained a spectral index of $\alpha=3.3\pm 0.13$ for
$S_\nu\propto \nu^\alpha$.
The spectral index gives $\beta=1.3\pm 0.13$ for the emissivity law, 
$\kappa_\nu\propto \nu^\beta$, using the approximated relation, 
$\beta=\alpha - 2$, which is valid for millimeter wavelengths 
\citep{2000prpl.conf..533B}.
Thus, given $\kappa_{\rm 230\,GHz}=0.01\,{\rm cm^2\,g^{-1}}$ 
\citep{1994coun.conf..179A}, we obtain a dust mass opacity of
$\kappa_{\rm 87\,GHz} = 2.8\times 10^{-3} \,{\rm cm^2\,g^{-1}}$.
The dust envelope mass is estimated to be 
$M_{\rm env}\approx 0.19\,M_\sun$ with a dust temperature of 
$30\,{\rm K}$ \citep{1993ApJ...414L..29C}.

\subsection{${\rm H^{13}CO^+}$ Line Emission}
\label{subsec:3.3}

\subsubsection{Integrated Intensity and Channel Maps}
\label{subsubsec:3.3.1}

\begin{figure}[htbp]
\includegraphics[width=.95\linewidth]{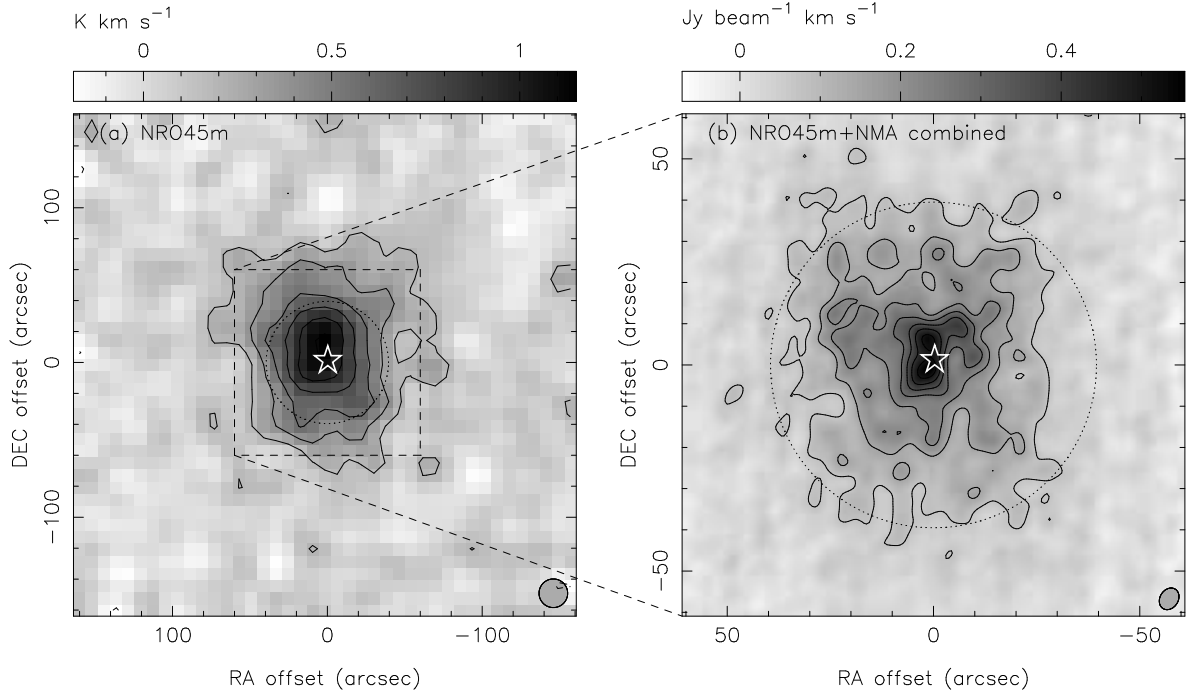}
\caption[
Total integrated intensity maps of B335 in the 
${\rm H^{13}CO^+} (J=1$--$0)$ line emission taken with the 45 m telescope.
]
{
Total integrated intensity maps of B335 in the ${\rm H^{13}CO^+} (J=1$--$0)$ 
line emission obtained with the 45 m telescope (left) and by 
combining the 45 m telescope and NMA data (right).
Contour intervals are $3\sigma$, starting from the $3\sigma$ levels of
the images where $1\sigma$ noise levels are $62\,{\rm mK\,km\,s^{-1}}$ in 
$T_{\rm mb}$ for the 45 m telescope map and 
$24\,{\rm mJy\,beam^{-1}\,km\,s^{-1}}$ for the combined map.
Dashed contours show the $-3\sigma$ level.
The open star in each map is the peak position of the $87\,{\rm GHz}$ 
continuum emission observed with the NMA.
The beam size for each map is shown as a filled circle or filled 
ellipse at the bottom right corner.
Dotted circle in the combined image indicates the field of view, i.e.,
FWHM primary beam size of the NMA observations.
}
\label{fig:3}
\end{figure}

The panels of Figure \ref{fig:3} show the integrated intensity images of 
the ${\rm H^{13}CO^+} (J=1$--$0)$ line emission over the LSR velocity range
from $8.08$ to $8.95\, {\rm km\,s^{-1}}$ made using the 45 m telescope 
data (left) and the combined 45 m telescope and NMA data (right).
We should note that this combined map includes the effect of the primary 
beam attenuation of the NMA so that extended emission toward the 
outside of the image seen in the 45 m telescope image is not reproduced in
the combined image.

The ${\rm H^{13}CO^+}$ emission in the 45 m telescope map has a 
single-peaked 
spatial distribution and shows an elongation from north to south.
The size of the ${\rm H^{13}CO^+}$ core above the $3\sigma$ level is 
$\sim 0.10\,{\rm pc}\times 0.09\,{\rm pc}$, with a P.A. of 
$\sim 0^\circ$.
The 45 m telescope plus NMA combined map shown in the right panel of 
Figure \ref{fig:3} clearly depicts the detailed structure of the 
inner core including large-scale flux distributions with a high resolution.
The higher contours of $\gtrsim 12\sigma$ show an elongated distribution 
from north to south with a size of 
$\sim 3000\,{\rm AU} \times1500\, {\rm AU}$, which is believed to be
an inner dense envelope associated with a central stellar source.
The envelope has a double peak near the center, and the 
$87\,{\rm GHz}$ continuum source is located between the peaks.
The elongation of the core and inner envelope is perpendicular to the 
molecular outflow axis \citep[e.g.,][]{1988ApJ...327L..69H}.
In the 45 m telescope map, there are faint ridges from the center along 
the P.A. of $\sim 70^\circ$, $\sim -45^\circ$, and $\sim -135^\circ$,
and they can be seen more clearly in the combined map.
Taking the outflow direction and opening angle into account 
\citep[$\sim 45^\circ$;][]{1988ApJ...327L..69H}, we believe that these 
ridges are related to the outflow activity.

\begin{figure}[htbp]
\includegraphics[width=.9\linewidth]{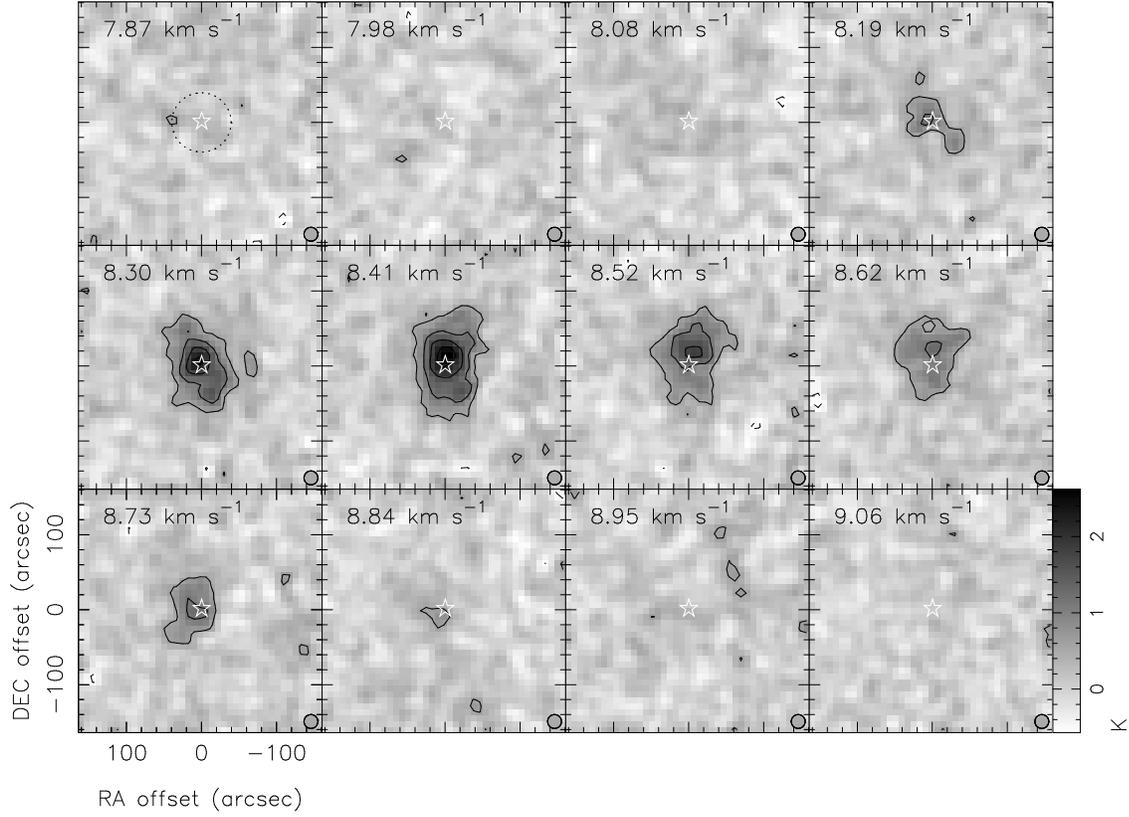}
\caption[
Velocity channel maps of B335 in the ${\rm H^{13}CO^+} (J=1$--$0)$ 
line emission taken with the 45 m telescope.
]
{
Velocity channel maps of B335 in the ${\rm H^{13}CO^+} (J=1$--$0)$ line 
emission taken with the 45 m telescope.
The central LSR velocity in ${\rm km\,s^{-1}}$ for each channel is shown 
in the upper left corner.
Contour intervals are the same as in Figure \ref{fig:3}, and the
$1\sigma$ noise level is $0.19\,{\rm K}$ in $T_{\rm mb}$.
The open star, filled circle in each panel, and dotted circle are the 
same as in Figure \ref{fig:3}.
}
\label{fig:4}
\end{figure}

\begin{figure}[htbp]
\includegraphics[width=.9\linewidth]{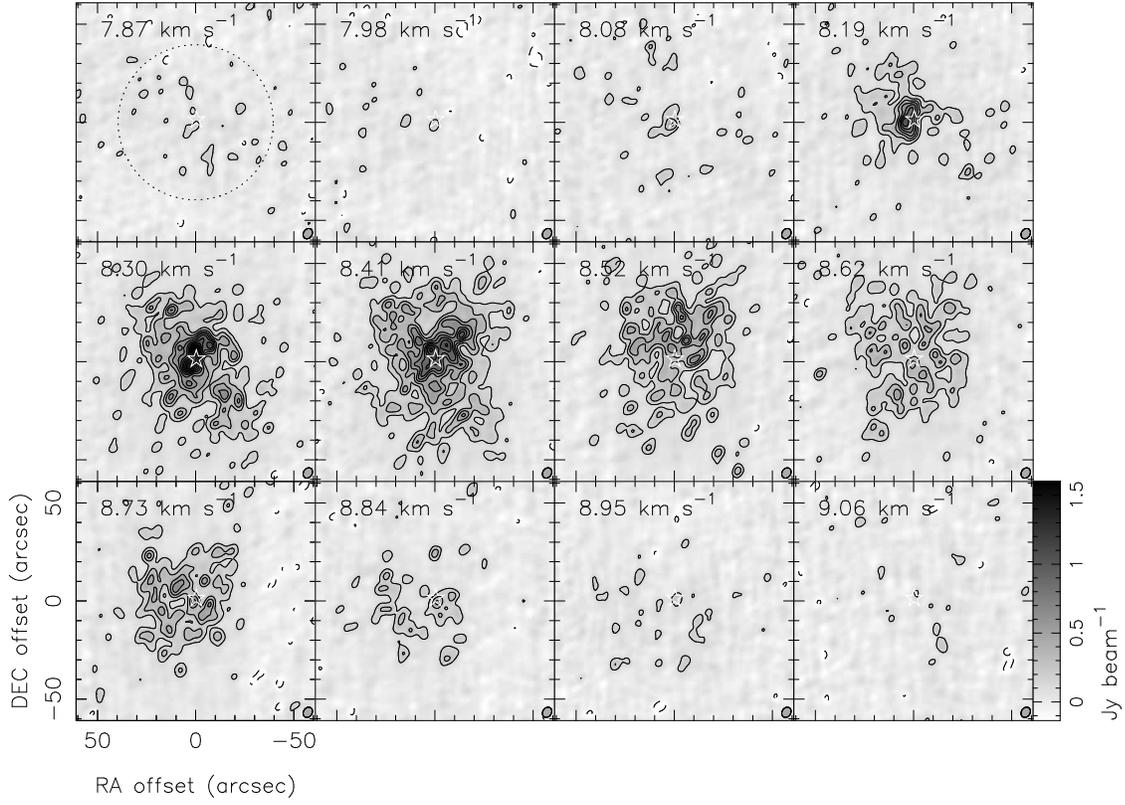}
\caption[
Velocity channel maps of B335 in the ${\rm H^{13}CO^+} (J=1$--$0)$
line emission obtained by combining the 45 m telescope and NMA data.
]
{
Velocity channel maps (CLEANed synthesized images) of B335 in the
${\rm H^{13}CO^+} (J=1$--$0)$ line emission obtained by combining the 45 m
telescope and NMA data.
The central LSR velocity in ${\rm km\,s^{-1}}$ for each channel is shown
in the upper left corner.
Contour intervals are the same as in Figure \ref{fig:3} and
$1\sigma$ noise level is $45\,{\rm mJy\,beam^{-1}}$.
The open star, filled ellipse in each panel, and dotted circle are the 
same as in Figure \ref{fig:3}.
}
\label{fig:5}
\end{figure}

In the velocity channel maps obtained with the 45 m telescope shown in 
Figure \ref{fig:4}, the emission peaks
are located at the north of the $87\,{\rm GHz}$ continuum source in the 
velocity range $V_{\rm LSR} = 8.19$--$8.62\,{\rm km\,s^{-1}}$, and at 
the south in the velocity range $V_{\rm LSR} = 8.73$--$8.84\,{\rm km\,s^{-1}}$.
This velocity gradient is perpendicular to the outflow axis and can be
interpreted as the rotational motion of the B335 core.
On the other hand, in the channel maps of combined data shown in 
Figure \ref{fig:5}, it is difficult to identify the corresponding velocity 
gradient because of the complicated emission distribution.
More detailed kinematics of the B335 core are discussed in Section 
\ref{subsubsec:3.3.3} and \ref{subsec:4.2} using position--velocity (PV)
diagrams made from the 45 m telescope and combined images.

\subsubsection{Mass and Column Density Profile in the Core}
\label{subsubsec:3.3.2}

In order to estimate the column density of the B335 protostellar core, we
analyzed the ${\rm H^{13}CO^+}$ images made from the 45-m telescope
and the combined NMA plus 45 m telescope data.

Under the local thermodynamic equilibrium (LTE) assumption, the 
${\rm H_2}$ column density can be calculated using the following 
formula:
\begin{eqnarray}
 N_{\rm H_2} &\simeq& 
 5.58 \times 10^{10}
 \frac{1}{
	 X({\rm H^{13}CO^+})
 }\nonumber\\
&& \times \frac{
	 T_{\rm ex}+0.69
 }{
	 e^{-4.16/T_{\rm ex}}}
 \frac{\tau}{1-e^{-\tau}}
 \left(
	 \frac{\int T_{\rm mb} d v}{\rm K\,km\,s^{-1}}
 \right)
  \quad{\rm cm^{-2}},
\label{eq:1}
\end{eqnarray}
where $T_{\rm ex}$ is the excitation temperature, $\tau$ is the 
optical depth of the ${\rm H^{13}CO^+}$ line, and $X({\rm H^{13}CO^+})$ 
is the fractional abundance of ${\rm H^{13}CO^+}$.
When deriving of Equation (\ref{eq:1}), we used the permanent dipole 
moment $\mu=4.07\,{\rm Debye}$ \citep{1979CPL....61..396H} and the 
rotational constant $B=43377.17\,{\rm MHz}$.
To obtain a complementary expression for the combined synthesized image, 
we convert the antenna temperature into the flux density $S_\nu$ in the
${\rm Jy\,beam^{-1}}$ through 
$S_\nu =\left(2k_{\rm B}/\lambda^2\right)\Omega_{\rm b}T_{\rm mb}$, where $\Omega_{\rm b}$ is the beam solid 
angle given by $\pi \theta_{\rm maj}\theta_{\rm min} /4\ln2$.
Thus, we have
\begin{eqnarray}
 N_{\rm H_2} &\simeq&
 9.01 \times 10^{12}
 \frac{1}{
	 X({\rm H^{13}CO^+})}
 \frac{T_{\rm ex}+0.69}{e^{-4.16/T_{\rm ex}}}
 \frac{\tau}{1-e^{-\tau}} \nonumber \\
&& \quad \times \left(
	\frac{\theta_{\rm maj}\theta_{\rm min}}{\rm arcsec^2}
 \right)^{-1}
 \left(\frac{\int S_\nu d v}{\rm Jy\,km\,s^{-1}}\right)
 \quad{\rm cm^{-2}}.
\label{eq:2}
\end{eqnarray}

We assumed the excitation temperature of ${\rm 15\,K}$ from the ${\rm
NH_3}$ analysis (Section \ref{subsec:3.1}) and the ${\rm H^{13}CO^+}$ 
line of the optically thin limit.
We assumed $X({\rm H^{13}CO^+})$ to be $8.3 \times 10^{-11}$ 
\citep{1987ApJ...313..320F}.
The derived total mass from the 45 m telescope map (the left panel of 
Figure \ref{fig:3}) is $\approx 1.2\,M_\sun$.

\begin{figure}[htbp]
\includegraphics[width=.9\linewidth]{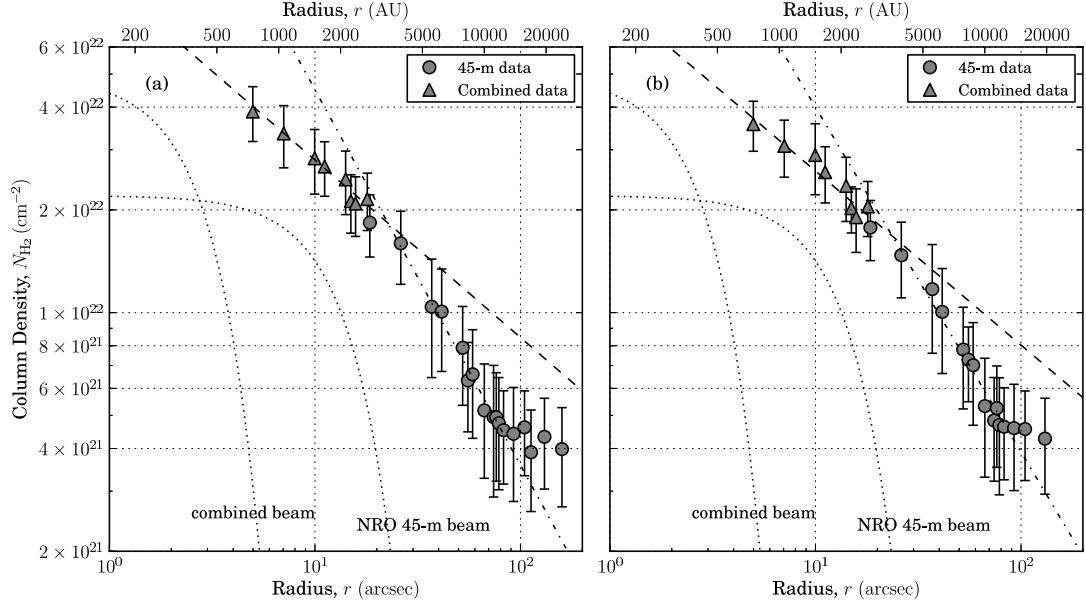}
\caption[
Radial profile of ${\rm H_2}$ column density in the B335 core.
]
{
Radial profile of ${\rm H_2}$ column density in the B335 core made from 
the ${\rm H^{13}CO^+}(J=1$--$0)$ image obtained with the 45 m telescope 
only (filled circles) and the combined 45 m telescope and NMA data 
(filled triangles).
Panel (a) shows the profile without the mask in the cavity 
regions and panel (b) shows that with the mask.
Dot-dashed and dashed lines in the plot indicate the best-fit results
of power-law fitting with the fitting range of $750$--$4000\,{\rm AU}$ and
 $4000$--$15{,}000\,{\rm AU}$, respectively.
Dotted curves indicate approximated beam patterns of 45 m telescope 
image and the combined image, Gaussian functions with the FWHM of 
$25\farcs0$ and $5\farcs0$, respectively.
}
\label{fig:6}
\end{figure}

We derived the radial column density profile in the B335 core from
the column density map calculated by the above formula.
The column densities were calculated from re-gridded images with cell 
sizes corresponding to the spatial resolutions of the images, i.e., $5''$ 
($750\,{\rm AU}$) for the combined image and $25''$ ($3750\,{\rm AU}$) for
the 45 m image.
The column density profile was made as a function of the radius from the 
$87\,{\rm GHz}$ peak using the distributions of the estimated column 
densities over the re-gridded cells.
In order to calculate the column density, the 45 m plus NMA image
needed to be corrected for the primary beam attenuation, which increases
the noise level in the outer region of the image.
Hence, the column density profile in the outer region of the core 
was derived using the 45 m telescope data, and the 45 m plus NMA combined 
data were used to fill in the inner region of $\lesssim 3000\,{\rm AU}$ 
where the data cannot be obtained from 45m telescope data.
Figure \ref{fig:6} represents the resulting ${\rm H_2}$ column density 
profiles obtained in the above procedure.
As suggested in Section \ref{subsubsec:3.3.1}, the B335 core could be 
affected by the outflow.
To examine this effect we also made the column density profile by 
masking out the regions with
P.A. of $67\fdg5$--$112\fdg5$ and $247\fdg5$--$292\fdg5$.
The masking angle of $45^\circ$ was chosen to match the opening 
angle of the outflow: $\sim 45^\circ$ by \citet{1988ApJ...327L..69H}
and $\sim 41^\circ$ by \citet{2001ApJ...563..903H}.
Figure \ref{fig:6}(a) and (b) show the ${\rm H_2}$ column density 
profile without and with the masking, respectively.
For both profiles, we can see that the column densities estimated from 
the 45 m and combined data are smoothly connected around the radius of 
$\sim 3000\,{\rm AU}$, and that the profiles in the inner radius are
shallower than those in the outer radius.

The column density profiles were fitted by two power-law 
functions of 
$N(r)=N_0(r/r_0)^{-s}$,
where $s$ is the power-law index and $N_0$ is the column density at 
$r = r_0 = 1500\,{\rm AU}$. 
Since Figure \ref{fig:6} clearly shows two different slopes between the 
inner and outer regions, we estimated the turnover radius at which the 
power-law index changes.
We evaluated the correlation coefficient of the power-law
fitting in the range of $r_{\rm fit,i}$--$r_{\rm fit,o}$, where
the inner radius ($r_{\rm fit,i}$) was variable and the outer radius 
($r_{\rm fit,o}$) was fixed to $r_{\rm fit,o} = 15{,}000\,{\rm AU}$.
As a result, we found that the correlation coefficient of the fitting 
decreased exceedingly when the fitting inner radius was
taken inside of $4000\,{\rm AU}$.
Thus, we adopted a turnover radius of $4000\,{\rm AU}$ in this paper
and performed the power-law fittings in the two regions of 
$750$--$4000\,{\rm AU}$ and $4000$--$15{,}000\,{\rm AU}$ separately.

For the inner region ranging from $750$ 
to $4000\,{\rm AU}$, we obtained 
$N_{\rm H_2}(r)=(2.8\pm1.0)\times 10^{22}(r/1500\,{\rm
AU})^{-(0.52\pm0.06)}~{\rm cm^{-2}}$ without masking, and 
$N_{\rm H_2}(r)=(2.6\pm1.0)\times 10^{22}(r/1500\,{\rm
AU})^{-(0.51\pm0.05)}~{\rm cm^{-2}}$ with masking.
For the outer region from $4000$ to $15{,}000\,{\rm AU}$, we obtained
$N_{\rm H_2}(r)=(4.5\pm1.3)\times 10^{22}(r/1500\,{\rm
AU})^{-(1.10\pm0.16)}~{\rm cm^{-2}}$ without masking and 
$N_{\rm H_2}(r)=(4.0\pm1.3)\times 10^{22}(r/1500\,{\rm
AU})^{-(1.01\pm0.16)}~{\rm cm^{-2}}$ with masking.
The column density profile obtained with masking was estimated to be 
slightly shallower than that without the masks.
From our fitting result, the column density at $r=6000\,{\rm AU}$ is 
$\approx 9.7\times 10^{21}\,{\rm cm^{-2}}$, which is consistent with 
the estimate by \citet{1999ApJ...518..334S}, 
$N_{\rm H_2}(6000\,{\rm AU})\sim6.3\times 10^{21}\,{\rm cm^{-2}}$.
Using our results, we examined the density distribution of the B335 core 
(discussed in Section \ref{subsubsec:4.1.1})
and estimated the total mass of gas associated with the B335 core 
to be $\approx 0.78M_\sun$ within the radius of $6000\,{\rm AU}$.

\subsubsection{Position--Velocity Diagrams}
\label{subsubsec:3.3.3}

\begin{figure}[htbp]
\includegraphics[width=.7\linewidth]{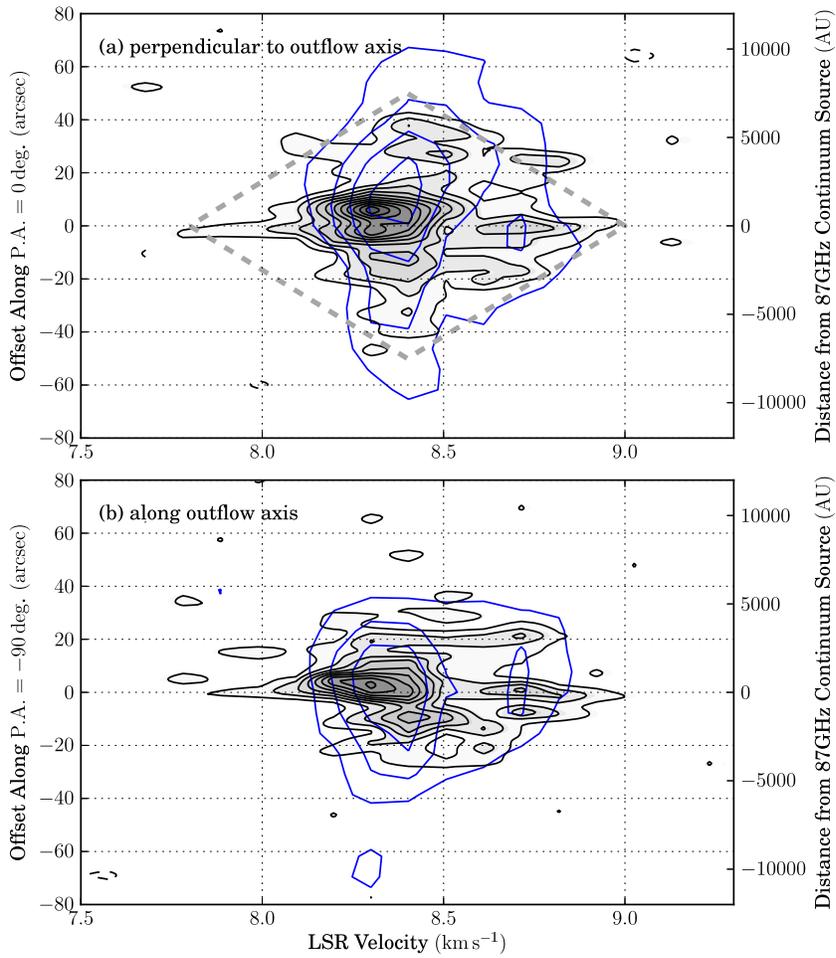}
\caption[
Position-velocity diagrams of the ${\rm H^{13}CO^+} (J=1$--$0)$ line 
emission for the B335 core.
]
{
Position-velocity diagrams of the ${\rm H^{13}CO^+} (J=1$--$0)$ line 
emission for the B335 core made from the 45 m telescope image data
(blue contours) and the combined data (gray scale with black contours).
Panel (a) and (b), respectively, show the 
diagrams along 
${\rm P.A.} =0^\circ$ (perpendicular to the outflow axis) and 
${\rm P.A.} =90^\circ$ (along the outflow axis), passing through the 
$87\,{\rm GHz}$ continuum source.
}
\label{fig:7}
\end{figure}

To examine the overall velocity structure in the B335 core, we made 
PV diagrams using the combined 
${\rm H^{13}CO^+}(J=1$--$0)$ line data as shown in Figure \ref{fig:7},
with black contours and gray scales.
We chose the two axes of ${\rm P.A.} = 0^\circ$ (perpendicular to the 
outflow axis) and ${\rm P.A.} = -90^\circ$ (along the outflow axis), 
passing through the $87\,{\rm GHz}$ continuum source.
The comparison with the diagrams made from the low-spatial-resolution 
45 m data overlaid with blue contours demonstrates a drastic effect of 
data combining.
Along the P.A. of $0^\circ$, the 45 m telescope data show a slight 
velocity gradient, $\sim 1.0\,{\rm km\,s^{-1}\,pc^{-1}}$ over a scale of
$20{,}000\,{\rm AU}$.
This gradient is comparable to the typical value of the cores in the 
Taurus dark cloud \citep{1993ApJ...406..528G}.
If the gradient represents a solid-body-like rotation of the core, 
then it is converted into an angular velocity of 
$\omega \sim 3.2\times 10^{-14}\,{\rm rad\, s^{-1}}$.
On the other hand, the diagram of P.A. of $-90^\circ$ does not show a 
conspicuous velocity gradient.

The diagrams of the 45 m and NMA combined data show line broadening
and a double-peaked intensity profile at the core center 
(offset $=0\arcsec$).
With a broadening line width toward the core center,
the diagram along the P.A. of $0^\circ$ shows an overall symmetrical 
diamond-like structure (represented by a dashed line in light gray).
We calculated intensity-weighted first and second moments defined by
$M_1 = \int I(v)vdv/\int I(v)dv$ and
$M_2=\sqrt{\int I(v)(v-M_1)^2dv/\int I(v)dv}$, respectively.
The second moment at the center (offset of $0\arcsec$) was 
$M_2 \approx 0.25\,{\rm km\,s^{-1}}$, whereas that at the offset 
of $20\arcsec$ was $0.17\,{\rm km\,s^{-1}}$.
The double-peaked profile at the core center is also slightly visible in 
the diagram made from the 45-m telescope data.
This profile has asymmetrical features that are more enhanced in the 
blueshifted peak than in the redshifted one.
This blue-skewed profile is maybe due to the infall 
motion of the core \citep{1993ApJ...404..232Z,1994ApJ...421..854Z}
because the ${\rm H^{13}CO^+}(J=1-0)$ line
emission is moderately optically thick around the systemic velocity
(see Section \ref{subsubsec:4.1.3}).

The center of the overall symmetrical emission distribution (lowest 
contour), as represented with a dashed diamond in Figure \ref{fig:7}(a),
is approximately $\sim 8.4\,{\rm km\,s^{-1}}$ and $0\arcsec$, 
which is considered to be the dynamical center of the B335 core.
Since the LSR velocity of $8.4\,{\rm km\,s^{-1}}$ is consistent 
with the peak velocities of the ${\rm NH_3}$ main components 
(Section \ref{subsec:3.1}), we adopt an LSR velocity of 
$8.4\,{\rm km\,s^{-1}}$ as the systemic velocity of the protostellar 
system.

\section{DISCUSSION}
\label{sec:4}

\subsection{Density Structure of the Core}
\label{subsec:4.1}

\subsubsection{Overall Structure of the Core}
\label{subsubsec:4.1.1}

When a molecular cloud core has a density distribution of $\rho(r)
\propto r^{-p}$, the profile of the column density integrated along the 
line of sight has a relation of $N(r)\propto r^{-s}\propto r^{1-p}$.
Hence, our obtained indices of $s \approx 0.5$ in the inner part of the 
B335 core and of $s \approx 1$ in the outer part correspond to 
$p\approx 1.5$ and $p\approx 2$, respectively.
We note that these indices barely varied even if we considered the 
influence of the cavity in the core made by the outflow.

The column density profile derived from our 45 m and combined data
with masking (Section \ref{subsubsec:3.3.2}) can be converted into the 
number density distribution:
\begin{eqnarray}
&&n(r) \approx 1.2\times 10^5 ~{\rm cm^{-3}}\nonumber\\
&& \left\{
\begin{array}{ll}
\left(\frac{r}{\rm 4000\,AU}\right)^{-1.51} \quad {\rm for~} 750\,{\rm AU}\leqslant r \leqslant 4000\,{\rm AU},\\
\left(\frac{r}{\rm 4000\,AU}\right)^{-2.01} \quad {\rm for~} 4000\,{\rm AU}\leqslant r \leqslant 15{,}000\,{\rm AU}.
\end{array}
\right.
\label{eq:3}
\end{eqnarray}

Most previous studies have derived density profiles for the B335 
star-forming core similar to our observational results: inner and outer 
indices of $-1.51$ and $-2.01$, respectively, with a 
turnover radius of $\sim 4000\,{\rm AU}$.
A power-law index of $p\sim 1.95$ for the density profile was estimated
from single-dish observations of the ${\rm C^{18}O}(J=1$--$0)$ and 
${\rm H^{13}CO^+}(J=1$--$0)$ line emissions (radius range of 
$4200$--$25{,}000\,{\rm AU}$) by \citet{1999ApJ...518..334S}.
From near-infrared extinction measurements,
\citet{2001ApJ...563..903H} suggested that the B335 core has a constant 
power-law index of $p= 1.91$ over the region $2000$--$15{,}000 \,{\rm AU}$, 
and otherwise displays inside-out collapse with an infalling radius 
(see Section \ref{subsec:4.2}) of $\sim 26\arcsec$ (corresponding to 
$\sim 3900\,{\rm AU}$). 
\citet{2003ApJ...583..809H,2003ApJ...596..383H} also suggested a single 
power law with $p= 1.55$ within the inner $\lesssim 3000 \,{\rm AU}$ region
from interferometric observations of $1.2$ and $3.0\,{\rm mm}$ 
continuum.
These results agree well with our estimate of the density profile from
the ${\rm H^{13}CO^+}(J=1$--$0)$ data.
Meanwhile, detailed analysis of SCUBA $450$ and $850\,{\rm \micron}$ 
continuum maps by \citet{2002ApJ...575..337S} showed that
the model with a power-law index of $p=1.8$ well describes the data 
as a best-fit result, and suggested an infalling radius of 
$1000\,{\rm AU}$, which is smaller than estimates in our as well as 
previous studies.
\citet{2010MNRAS.406.1190D} recently conducted an 
unbiased fitting to the dust continuum observations toward B335.
For the power-law density distribution, they obtained
$p=1.5$--$1.9$ throughout the envelope, although they did not find strong 
evidence of inside-out collapse with an infalling radius of 
$\sim1000\,{\rm AU}$.

\subsubsection{Inner Structure of the Core}
\label{subsubsec:4.1.2}

We confirmed the above estimate of the density profile from our combined 
${\rm H^{13}CO^+}(J=1$--$0)$ line data using an analysis of the visibility 
data of the $87\,{\rm GHz}$ continuum emission observed with the NMA.
This approach of directly examining in the $u$--$v$ domain enables us 
to avoid possible artifacts caused by the deconvolution of interferometric 
images.

When discussing the emission distribution of the inner region of an 
envelope observed with interferometers, we should consider the temperature 
distribution of the envelope.
We consider the observed intensity as a function of the impact parameter, 
$I_\nu(r_{\rm b})$, from an optically 
thin dust envelope that has a spherical density distribution, 
$\rho(r)$, and a dust temperature, $T_{\rm d}(r)$.
If the density and temperature follow radial power laws, 
$\rho\propto r^{-p}$ and $T_{\rm d}\propto r^{-q}$, and if we assume 
that the opacity does not vary along the line of sight, then the intensity 
profile also has a power-law profile, 
$I_\nu (r_{\rm b})\propto r_{\rm b}^{-(p+q-1)}$, in the 
Rayleigh--Jeans regime.
We assumed that the extent of source intensity $I_\nu(r_{\rm b})$ is 
sufficiently compact in the primary beam.
The visibility amplitude as a function of $u$--$v$ distance ($b$) can be 
given by $V(b) \propto b^{p+q-3}$ for 
$3/2 < p+q < 3$ (see the Appendix for details). 

Figure \ref{fig:8} shows a plot of binned visibility 
amplitudes as a function of the $u$--$v$ distance for the $87\,{\rm GHz}$ 
continuum data.
The binning is logarithmic, and the amplitudes are obtained as the 
vectorial average of the complex visibilities in each bin.
The decrease in visibility amplitude with $u$--$v$ distance can be
interpreted as a power law.
We fitted the visibility amplitude profile with a power-law function of 
the form $A\times b^{-\xi}$.
The best-fit power-law index is $\xi=1.13\pm0.06$ (solid line in the plot),
which can be converted into a power-law index with an intensity 
distribution of $p+q-1 = 0.87$.
The power-law index of dust temperature distribution is expected to 
depend on the opacity, $T_{\rm d}(r)\propto L^{q/2}r^{-q}$, where 
$q=2/(4+\beta)$ \citep{1994ApJ...424..729D}.
Thus, the power-law index of the density distribution is estimated to be
$p=1.49$ for $\beta = 1.3$ (see Section \ref{subsec:3.2}), which shows a 
good agreement with the estimate from the combined ${\rm H^{13}CO^+}$ data 
(Section \ref{subsubsec:4.1.1}).

\subsubsection{Uncertainties of Density Profile}
\label{subsubsec:4.1.3}

There are several uncertainties included in the derived 
density distribution.
In this section, we examine the effect of opacity, assumption of 
uniform temperature, index conversion from column density to volume 
density profiles, and ${\rm H^{13}CO^+}$ fractional abundance.

First, we investigate the effect of the optical depth of 
${\rm H^{13}CO^+}$ line emission, because the decrement of derived 
column densities in the inner radii might be due to larger optical depth 
within the $5\arcsec$ beam.
\citet{1999ApJ...518..334S} indicated that the ${\rm H^{13}CO^+}$
emission is optically thin even at the center of an $18\arcsec$ beam 
from the measured ${\rm HC^{18}O^+}$ to ${\rm H^{13}CO^+}$ 
peak intensity ratio of $\sim 5.8$ and expected abundance ratio, 
$X({\rm H^{13}CO^+})/X({\rm HC^{18}O^+})$, of $5.5$. 
Furthermore, the integrated intensity ratio 
of ${\rm HC^{18}O^+}$ to ${\rm H^{13}CO^+}$ implies that ${\rm H^{13}CO^+}$
emission should be optically thin at the center. 
According to Figure 1 in \citet{1999ApJ...518..334S}, the line ratio 
actually ranges from $\sim2.8$ to $\sim5.8$ across the velocity;
the lowest line ratio indicates that ${\rm H^{13}CO^+}$ emission is 
marginally thick, $\tau \sim 2$ at maximum in certain velocities.
This opacity enhancement is limited only to a narrow velocity range, so 
it does not introduce a significant error when estimating the column 
density near the center, i.e., the underestimate. 
Therefore, the effect of the optical depth cannot account for the 
turnover around the radius of $4000\, {\rm AU}$.

Second, we assumed that the core has a uniform temperature distribution
that does not have much effect on the estimate of the column density.
If we adopt the temperature profile from \citet{2005ApJ...626..919E}, 
for example, then the difference in temperature between radii of 
$750\,{\rm AU}$ and $20{,}000\,{\rm AU}$ should be a factor of $\sim 2$.
From Equation (\ref{eq:2}), the uncertainty of the column density 
estimate expected by this temperature variation could be derived as
$\pm \sim10\%$ at most.

Third, the derivation of the density profile of the core from the column 
density distribution depends on the radial finiteness of the core traced
by the ${\rm H^{13}CO^+}$ molecular line emission.
For a spherical core with a size of $R_{\rm out}$ and a power-law density 
profile of $\rho (r)=\rho_0(r/r_0)^{-p}$,
the column density that is given by integrating the densities along the 
line of sight at the impact parameter $r_{\rm b}$ can be expressed as
\begin{equation}
 N(r_{\rm b}) = \int^{\arccos(r_{\rm b}/R_{\rm out})}_{-\arccos(r_{\rm b}/R_{\rm out})} \rho (r) d\theta = N_0 \left(\frac{r_{\rm b}}{r_0}\right)^{-p+1},
 \label{eq:4}
\end{equation}
where
\begin{equation}
 N_0 = \rho_0 r_0 
 \int^{\arccos(r_{\rm b}/R_{\rm out})}_{-\arccos(r_{\rm b}/R_{\rm out})} 
 \cos^p\theta d\theta,
\label{eq:5}
\end{equation}
and the variable $\theta$ is an angle between a radial vector \mbox{\boldmath $r$} and
the plane of the sky.
Therefore, the coefficient of $N_0$ is also a function of $r_{\rm b}$ so 
that the column density profile deviates from a power-law of the form 
$N(r_{\rm b}) \propto r_{\rm b}^{-p+1}$.
Moreover, if the core has different power-law dependencies between the 
inner and outer regions, then the simple prediction of the column density 
profile in Equation (\ref{eq:4}) is not exactly valid.
Figure \ref{fig:9} shows the column density profiles of cores that have
power-law density distributions with a cutoff radius $R_{\rm out}$.
The dashed lines indicate the profiles for the predictions of
$\propto r^{-p+1}$.
For the case of a single power-law density distribution of 
$\rho (r)\propto r^{-2}$ as shown in Figure \ref{fig:9}(a), the resulting
column density coincides well with the dependency of $\propto r^{-1}$ at 
radii smaller than $\sim 0.1R_{\rm out}$.
However, it tends to depart from the power-law dependency and become
steeper with increasing radius.
Figure \ref{fig:9}(b) shows the column density 
profile for the core with density distributions of $\rho (r)\propto 
r^{-1.5}$ for $r\leqslant 0.125 R_{\rm out}$ and $\rho (r)\propto r^{-2}$ 
for $r> 0.125R_{\rm out}$.
This plot demonstrates that the column density profile of a core that 
has inner and outer regions with different power-law density 
distributions can be estimated to be steeper than the simple prediction of 
$N(r_{\rm b}) \propto r_{\rm b}^{-p+1}$.
We examined the uncertainties caused by these effects when converting the 
column density profile into the density profile, and we found that the 
true density profile is likely to be shallower by $\sim 0.1$ at 
most in the power-law index than for our estimates.

Finally, as for the abundance of ${\rm H^{13}CO^+}$, 
recent studies of chemical evolution in star-forming cores have shown
that the abundance of ${\rm HCO^+}$, which is a daughter species of 
${\rm CO}$, decreases at radii that are smaller than 
the ${\rm CO}$ sublimation radius 
\citep{2004ApJ...617..360L,2008ApJ...674..984A}.
Nevertheless, it is shown that the abundance of ${\rm HCO^+}$ at the 
inner radii increases with the evolution of core collapse after a 
central stellar object is born.
\citet{2005ApJ...626..919E} simulated a large number of molecular
line profiles from B335 using various physical models. 
By the use of a self-consistent chemical model with core evolution, the 
result showed that the abundance of ${\rm HCO^+}$ hardly changes along the 
core radius.
The shallower profile at smaller radii can be affected by the
radial distribution of the ${\rm H^{13}CO^+}$ abundance, although it is 
not expected to be dominant.

The possible effects of radial dependence in the ${\rm H^{13}CO^+}$ 
abundance and the finite core radius oppose one another.
It is difficult to estimate how they contribute to our data.
Nevertheless, by taking into account the agreement of the column density 
profiles in the inner region of the B335 core between the derivations from
the combined ${\rm H^{13}CO^+}$ and dust continuum data, we determine that
the effects of variation in optical depth and radius dependence in the 
${\rm H^{13}CO^+}$ abundance are likely to be negligible.
Moreover, the agreement between the profile indices derived from our data 
with those from the extinction analysis of near-infrared data indicates 
that the effect of these uncertainties, especially in the temperature and 
fractional abundance of ${\rm H^{13}CO^+}$, is not very considerable.
In other words, our ${\rm H^{13}CO^+}$ data well represent the column 
density structure of the core, which mostly ensures the validity of our 
analysis for the physics of star formation.

\subsection{Velocity Structure of the Core}
\label{subsec:4.2}

We performed model calculations of the PV diagrams and 
investigated which model reproduces the observed signatures of 
the PV diagram in the ${\rm H^{13}CO^+}$ line well.
For comparison with the observed results, the model calculations were 
performed with different radial distributions of the infalling velocities 
and also for three different assumed central stellar masses.
    
\subsubsection{Model Calculations of PV Diagram}
\label{subsec:4.2.1}

We assume a contracting spherical star-forming core
with rotation as described in detail below.
Most of the parameters for the calculations simulating PV 
diagrams were taken from our observational results and previous 
studies.

The core has a size of $20{,}000\,{\rm AU}$ and the radial density 
profile in Equation (\ref{eq:3}) shows power-law dependencies 
($\rho\propto r^{-p}$) of 
$p = 1.5$ in the inner ($r < 4000\,{\rm AU}$) and $p = 2$ in the outer 
($r \geqslant 4000\,{\rm AU}$) regions.
Such a difference in power-law indices is naturally expected from the 
isothermal collapse model because of the boundary of the inner 
free-falling region and the outer region in which the condition at the 
stage of the formation of a central stellar object is expected to be 
conserved.
The boundary radius is referred to as the infalling radius ($r_{\rm inf}$),
and therefore our derived density profile indicates 
$r_{\rm inf} = 4000\,{\rm AU}$.

We also introduced the rotational motion of the core.
The inclination angle of the rotation axis from the plane 
of the sky should be considered because it affects the 
line-of-sight velocity of rotational motion.
We assumed that the rotation axis corresponds to the outflow axis
whose inclination angle for B335 was estimated to be $10^\circ$ 
\citep{1988ApJ...327L..69H}, $8\pm5^\circ$ 
\citep{1988ApJ...334..196C}, and $9\pm1^\circ$ \citep{1989ApJ...338..952M}.
We adopted an inclination angle of $10^\circ$ for the rotation axis in 
our calculations.
In the free-fall region with $r \leqslant r_{\rm inf}$, we adopted the 
increasing rotational velocity which is inversely 
proportional to the radius owing to the angular momentum conservation
during collapse.
Such a velocity field in infalling envelopes was indeed suggested 
observationally \citep[e.g., L1551-IRS5 by][]{1998ApJ...504..314M}. 
The overall velocity gradient from the 45 m telescope observations,
$1.0\,{\rm km\,s^{-1}\,pc^{-1}}$ ($\omega = 3.2\times 10^{-14}\,{\rm rad\, s^{-1}}$), 
can be regarded as the initial angular momentum of the core, 
which is still preserved in the outer radius of 
$r \geqslant r_{\rm inf} = 4000\,{\rm AU}$.
For the thermal line broadening, we set the line-of-sight velocity width 
of $0.15\,{\rm km\,s^{-1}}$ for ${\rm H^{13}CO^+}$ molecules at 
$T_{\rm k} = 15\,{\rm K}$.

We calculated radiative transfer equations to 
estimate the relative intensity distributions and verify the effect of 
optical depth.
Assuming a two-level state, the populations in rotational levels of an
${\rm H^{13}CO^+}$ molecule were 
calculated using the Einstein $A$ coefficient and the collisional rate 
coefficients with ${\rm H_2}$ from 
\citet{2009A&A...494..977G} and \citet{1999MNRAS.305..651F}.
The final model PV diagrams were smoothed with Gaussian functions 
whose FWHMs are the same as actual resolutions of the combined image
for both the spatial and velocity directions.

\subsubsection{Comparison of Model Calculations of PV Diagrams with Observations}
\label{subsec:4.2.2}

We conducted simulations by varying the parameters
of the inward velocity in the outer region ($r\geqslant r_{\rm inf}$) of 
the core, $v_{r\geqslant r_{\rm inf}}$, and the mass of the central 
stellar object to compare with observational results.
We adopted the inward velocities in the outer region of 
$v_{r\geqslant r_{\rm inf}}= 0\,{\rm km\,s^{-1}}$, 
$c_{\rm s}$, and 
$3.3\,c_{\rm s}$.
The central stellar mass ($M_{\rm c}$) for B335 was estimated in a wide 
range in previous studies, e.g.,  
\citet{1995ApJ...448..742C} estimated it to be $0.37\,M_\sun$ while
$0.04M_\sun$ was suggested
by \citet{2010ApJ...710.1786Y}. 
In our model calculations, we used three values of the central stellar 
mass: $M_{\rm c}=0.05M_\sun$, $0.1M_\sun$, and $0.15M_\sun$.
Figure \ref{fig:10} shows the calculated PV
diagrams perpendicular to the rotation axis passing through the core 
center.

As for the inward velocities in the outer region of the core, 
the calculated PV diagrams for 
$v_{r\geqslant r_{\rm inf}}= 0\,{\rm km\,s^{-1}}$ well 
reproduce the features in the observed PV diagram.
The inward velocity in the outer region affects the overall line width 
in the PV diagram.
Although the PV diagrams calculated with 
$v_{r\geqslant r_{\rm inf}}=c_{\rm s}$ are also similar to the observed result, 
the overall velocity widths seem to be larger than that of the observed 
one. The simulated diagrams with 
$v_{r\geqslant r_{\rm inf}}=3.3\,c_{\rm s}$ undoubtedly disagree with the 
observation.
We found that the central stellar mass mainly contributes
line broadening at the core center, which means that the infalling motion 
is more dominant than the spin-up rotation owing to the conservation of 
angular momentum around the core center.
For the diagrams with $v_{r\geqslant r_{\rm inf}}= 0\,{\rm km\,s^{-1}}$, 
it seems that line wings at the center position are not enough in the case
of $M_{\rm c}=0.05M_\sun$ while they are excessive in the case of
$M_{\rm c}=0.15M_\sun$.
Consequently, we determine in our model calculations that the PV diagram
with $v_{r\geqslant r_{\rm inf}}= 0\,{\rm km\,s^{-1}}$ and 
$M_{\rm c}=0.1M_\sun$ well represents the observed PV diagram.

The calculated model PV diagram with 
$v_{r\geqslant r_{\rm inf}}= 0\,{\rm km\,s^{-1}}$ and $M_{\rm c}=0.1M_\sun$ 
shows that the peak is skewed toward the blueshifted velocity and has a 
shoulder in the 
profile around the systemic velocity at the center position.
These features in the model diagram are due to the ${\rm H^{13}CO^+}$ 
optical depth and are similar to the observational result.
Hence, the suggestion discussed in Section \ref{subsubsec:4.1.3} that the 
${\rm H^{13}CO^+} (J=1$--$0)$ line is marginally optically thick is 
probably reasonable.

In addition, we can obtain the radial distribution of the specific angular 
momentum from the profile of rotation velocity in the model calculations 
of PV diagrams.
Figure \ref{fig:11} shows the radial distribution of specific angular 
momentum in the model that well represents the observation:
the case of
$v_{r\geqslant r_{\rm inf}}= 0\,{\rm km\,s^{-1}}$ and 
$M_{\rm c}=0.1M_\sun$.
We can see the spin-up rotation within the infall radius of 
$r_{\rm inf} = 4000\,{\rm AU}$ as deviations 
from the solid-body rotation with an angular velocity of 
$\omega = 3.2\times 10^{-14}\,{\rm rad\,s^{-1}}$.
In previous observational studies, the specific angular momenta of the 
core and envelope in B335 have been measured at several radii 
from velocity gradients across the outflow axis.
\citet{2010ApJ...710.1786Y,2011ApJ...742...57Y} derived the specific 
angular momenta of $\lesssim 7\times10^{-5}\,{\rm km\,s^{-1}\,pc}$ at a 
radius of $370\,{\rm AU}$ and $8\times10^{-5}\,{\rm km\,s^{-1}\,pc}$ at
$90\,{\rm AU}$ from velocity gradients seen in the ${\rm C^{18}O}(J=2$--$1)$ 
and ${\rm CS}(J=7$--$6)$ envelopes, respectively.
They discussed the evolution of specific angular momentum,
including the measurements by \citet{1999ApJ...518..334S} of
$\sim 5.4\times 10^{-4}\,{\rm km\,s^{-1}\,pc}$ at $1000\,{\rm AU}$ and
$\sim 4.6\times 10^{-3}\,{\rm km\,s^{-1}\,pc}$ at $20{,}000\,{\rm AU}$.
These measurements are also indicated in Figure 
\ref{fig:11} as open squares and open triangles. 
This plot shows that our model calculations successfully derived the 
radial distribution of the specific angular momentum consistent with 
measurements from previous observational studies.
In particular, specific angular momenta that were recently estimated
from high-resolution observations in submillimeter wavelengths at
radii of $90\,{\rm AU}$ and $370\,{\rm AU}$ show good agreement with the
results of our model calculations.

\subsection{Comparison with Theoretical Models}
\label{subsec:4.3}
The density and velocity structures of a collapsing molecular cloud 
core are crucial to distinguish theoretical models of gravitational 
collapse.
We examine the observational results by comparing them 
with the key properties of the two star formation models, i.e., 
the \citet{1977ApJ...214..488S} and 
Larson--Penston solutions 
\citep[][hereafter 
``LP solution'']{1969MNRAS.145..271L,1969MNRAS.144..425P}.
Although the Shu and LP pictures of core contraction are similar in terms 
of the power-law dependencies of density distribution and inward motion,
it is possible to distinguish between the two models by quantitative 
comparisons.

The power-law dependency of the density profile in the B335 core 
qualitatively matches the similarity solutions for an isothermal 
spherical cloud in its post-protostar formation stage
\citep{1977ApJ...214..488S,1977ApJ...218..834H}, which has also been 
demonstrated by numerical simulations
\citep[e.g.,][]{1999PASJ...51..637O}; i.e., $\rho \propto 
r^{-3/2}$ in the dynamical free-fall region and 
$\rho \propto r^{-2}$ 
in the outer region where the condition at the protostar formation 
stage should be conserved. 
The absolute density of the core is one of the key parameters used to 
discriminate between isothermal collapse models.
We derived the number density of the B335 core as $\approx 1.2\times 
10^5\,{\rm cm^{-3}}$ at the radius of $4000\,{\rm AU}$.
In Shu's inside-out picture, the density distribution in the outer 
region of the core is expected to correspond to the singular 
isothermal sphere, $\rho(r)=c_{\rm s}^2/2\pi G\,r^{-2}$.
From this model, we obtain 
$n(r={\rm 4000\,AU})=9.2\times 10^4\,{\rm cm^{-3}}$
with $T=15\,{\rm K}$.
On the other hand, the ``runaway collapse'' of the LP solution during
the core formation has a density that is $4.43$ times higher than that in 
Shu's solution, so that we obtain $4.1\times 10^5\,{\rm cm^{-3}}$ at 
$r=4000\,{\rm AU}$.
Our derived density is comparable to that predicted by Shu's solution 
with a factor of $\approx 1.4$, however, it is considerably smaller than 
that of the LP solution.

The inward velocity in the outer region from our model calculations also 
supports Shu's solution.
As described above our model calculations of PV diagrams successfully 
explain the observed features within the uncertainties, in which the mass 
of the central stellar source and the inward velocity in the outer core 
region are estimated.
Here, we focus on the inward motion in the outer region, which is one of 
the key characteristics to
distinguish the gravitational collapse models.
The static outer region of the core
($v_{r\geqslant r_{\rm inf}}=0\,{\rm km\,s^{-1}}$) represents the velocity 
structure of Shu's solution.
On the other hand, numerical simulations of isothermal collapse
models, which have been conducted to compare with similarity solutions, 
have showed that the inward velocity in the runaway collapse phase is not 
a constant but a function of age and radius.
These inward velocities can be observed as intermediate pictures
between those of the LP and Shu models, and are important characteristics 
of the collapse of an isothermal sphere.
The infall velocity in the runaway collapse phase is determined by the 
initial ratio of gravitational force to pressure force 
\citep{1999PASJ...51..637O}.
The cloud core that begins to collapse from a marginally unstable 
isothermal gas sphere has a subsonic inward velocity in the outer region.
Even from our model comparisons with the PV diagram of the observed data, 
we cannot exactly clarify if the outer region of the core has a subsonic 
inward velocity. 
However, it is plausible that there is an inward velocity 
$\gtrsim c_{\rm s}$ in the outer region, such as 
$v_{r\geqslant r_{\rm inf}}=3.3\,c_{\rm s}$ which 
represents LP solution, can be ruled out.

Consequently, we suggest that the B335 core has initiated its 
gravitational collapse from a quasi-static initial condition similar to
Shu's model.
Otherwise, it is also possible to explain our observed results by
means of an isothermal collapse of a cloud core that has a mass slightly
larger than the Bonner--Ebert mass
\citep{1993ApJ...416..303F,1999PASJ...51..637O}.
In Shu's similarity solutions, an infalling radius provides a rough 
age of the cloud after a point source is formed at the core center, 
because the boundary between the infalling inner region and the outer 
region propagates outward as a rarefaction wave with a velocity of the
isothermal sound speed.  
In our case, we take the turnover radius to be the infalling radius and
obtain an age of $\sim 8\times 10^4\,{\rm years}$ which is comparable 
to the order of the age of the Class 0 phase.
Adopting the mass infall rate predicted by Shu's solution, 
$\dot{M}=0.975 c_{\rm s}^3/G$, along with the estimated age, we obtain a
central stellar mass of $\approx 0.2M_\odot$.
The estimate of the central stellar mass from our model calculation is 
closer to the prediction from Shu's solution than that from the LP model,
which expects a 48 times higher mass infall rate.
Therefore, the above comparisons between our results and the theoretical 
models indicate that the picture of Shu's model is more preferable than 
that of the LP solution for the B335 core.

\section{SUMMARY}
\label{sec:5}

We presented a study of the dense molecular cloud core harboring the
low-mass protostar, B335 in the ${\rm H^{13}CO^+}
(J=1-0)$ molecular line emission using the Nobeyama 45 m telescope and the 
NMA. 
Our main findings are summarized as follows.

1. The single-dish observations revealed a dense core with a size of
$\sim 0.10\,{\rm pc}\times 0.09\,{\rm pc}$.
Our analysis using a combining technique of single-dish and interferometer 
data revealed the structure of the inner dense envelope within the core
with a high spatial resolution of $\sim 750\,{\rm AU}$. 
The envelope size is 
$\sim 3000\,{\rm AU}\times 1500\,{\rm AU}$.
Both of them have an elongated distribution toward the north--south 
direction, perpendicular to the outflow axis.
The mass of the core is estimated to be $\approx 1.2M_\sun$

2. We determined the radial column density profile of the B335 core and 
found a
reliable difference between the power-law indices of the outer 
and inner regions of the dense core.
The turnover radius is considered to be $\sim 4000\,{\rm AU}$, which is 
consistent with the infalling radius estimated in previous work. 
Our derived density profile, 
$n(r) \approx 1.2 \times 10^5 \left(r/{\rm 4000\,AU}\right)^{-1.51}~{\rm cm^{-3}
}$ 
for
$750\,{\rm AU} \leqslant r \leqslant 4000\,{\rm AU}$ and 
$n(r) \approx 1.2 \times 10^5 \left(r/{\rm 4000\,AU}\right)^{-2.01}~{\rm cm^{-3}
}$ 
for
$4000\,{\rm AU} \leqslant r \leqslant 15{,}000\,{\rm AU}$, is better explained,
both qualitatively and quantitatively, in the picture of Shu's 
self-similar solution than in that of the LP solution.

3. The dense core shows a slight overall velocity gradient of 
$\sim 1.0\,{\rm km\,s^{-1}}$ over the scale of $20{,}000\,{\rm AU}$ 
across the outflow axis.
This velocity gradient is considered to represent a solid-body 
rotation and corresponds to an angular velocity of 
$\omega \sim 3.2\times 10^{-14}\,{\rm rad\,s^{-1}}$.
Our combined image also revealed detailed velocity structures in 
the dense core with a high resolution.
The velocity structure of the B335 core can be well explained in 
terms of the collapse of an isothermal sphere, 
in which the core has an inner free-fall region and an outer 
region preserving the condition at the stage of protostar formation.

4. We performed simple model calculations of PV diagrams 
to examine the observed diagrams.
The model calculations successfully reproduce observational results,
while suggesting a central stellar mass of $\sim0.1\,M_\odot$ and a small 
inward velocity of $\sim 0\,{\rm km\,s^{-1}}$ in the outer region of 
the core $\gtrsim 4000\,{\rm AU}$.

5. Quantitative comparisons of density and velocity structures from
the observational results with theoretical models
show an agreement with Shu's quasi-static inside-out star formation.
Furthermore, it is possible for the outer region of the B335 core to have 
a subsonic inward velocity.
We concluded that a picture of Shu's solution or an isothermal 
collapse of a marginally stable Bonnor--Ebert sphere is suitable for the 
gravitational collapse of the B335 core.

\acknowledgments
This study was based on observations at the Nobeyama Radio Observatory 
(NRO), which is a branch of the National Astronomical Observatory of 
Japan, National Institutes of Natural Sciences.
The authors are grateful to the staff at the NRO for operating the NMA 
and the 45 m telescope, helping us with the data reduction.
We also thank an anonymous referee whose comments significantly improved 
the paper. Y.K. thanks Ken'ichi Tatematsu and the staff at 
the NAOJ Chile Observatory for their helpful comments, continuing 
interest, and encouragement.

\appendix
\section{ANALYSIS FOR THE VISIBILITY FUNCTION OF DUST ENVELOPE}
\label{app:a}
In Section \ref{subsubsec:4.1.2}, we discuss the emission distribution 
of the inner region of the envelope using interferometric data in 
the $u$-$v$ domain.
The detailed expression of the analysis is described in this paper.

For optically thin dust emission, the observed intensity from an envelope
that has a spherical density distribution, $\rho(r)$, and a dust 
temperature, $T_{\rm d}(r)$, as a function of the impact parameter,
$r_{\rm b}$, is written as
\begin{equation}
 I_\nu(r_{\rm b}) = 2\int^{r_{\rm out}}_{r_{\rm b}} B_\nu \left(T_{\rm d}(r)\right)\kappa_\nu(r)\rho(r) \frac{r}{\sqrt{r^2 - r_{\rm b}^2}} dr,
\label{eq:b1}
\end{equation}
where $r_{\rm out}$ is the outer radius of the envelope.
If the density and temperature follow radial dependencies of power-laws, 
$\rho\propto r^{-p}$ and $T_{\rm d}\propto r^{-q}$, and if we assume that 
the opacity does not vary along the line of sight, then the intensity 
also has a power-law profile, 
$I_\nu (r_{\rm b})\propto r_{\rm b}^{-(p+q-1)}$, in 
the Rayleigh--Jeans regime.
We assume that the intensity distribution, $I_\nu(r_{\rm b})$, is more 
compact in extent than the primary beam of interferometric observations, 
and gain variations during the observations are properly corrected.
The visibility as a function of $u$--$v$ distance, 
$b=\left(u^2+v^2\right)^{1/2}$, can be given by the 
Hankel transform of the intensity distribution,
\begin{equation}
 V(b) = 2\pi \int^\infty_0 I_\nu(r_{\rm b})J_0\left(2\pi r_{\rm b}b\right)
 r_{\rm b}d r_{\rm b},
\label{eq:b2}
\end{equation}
where $J_0(z)$ is a zeroth-order Bessel function.
 
Equation (\ref{eq:b2}) is rewritten as a function
of $u$--$v$ distance $b$,
\begin{equation}
 V(b)=\int^\infty_0 I_\nu(r_{\rm b}) \int^{2\pi}_0 
\exp\left[-2\pi i b r_{\rm b} \cos(\theta - \alpha)\right] 
d\theta r_{\rm b} d r_{\rm b},
\label{eq:b3}
\end{equation}
where $(l,m)=(r_{\rm b}\cos\theta, r_{\rm b}\sin\theta)$ and 
$(u,v)=(b\cos\alpha, 
b\sin\alpha)$.
By definition, a zeroth-order Bessel function is given by
\begin{equation}
 J_0(z) = \frac{1}{2\pi}\int^\infty_0 \exp\left(-iz\cos\theta\right),
\label{eq:b4}
\end{equation}
so we obtain
\begin{equation}
 V(b) = 2\pi \int^\infty_0 I_\nu(r_{\rm b})J_0\left(2\pi r_{\rm b}b\right)
 r_{\rm b} d r_{\rm b}.
\label{eq:b5}
\end{equation}
This is the Hankel transform of the intensity distribution.
We expect the intensity distribution to have a power-law dependency,
this integral has a solution of the form \citep{1994tisp.book.....G}
\begin{equation}
 \int^\infty_0 x^\mu J_0(ax)dx = 2^\mu a^{-\mu-1}
\frac{\Gamma\left(\frac{1}{2}+\frac{1}{2}\mu\right)}
{\Gamma\left(\frac{1}{2}-\frac{1}{2}\mu\right)},
\label{eq:b6}
\end{equation}
for
\begin{equation}
-1 < \mu < \frac{1}{2}, \quad a > 0,
\label{eq:b7}
\end{equation}
where $\Gamma (z)$ is the Gamma function:
\begin{equation}
\Gamma (z) = \int^\infty_0 e^{-t} t^{z-1} dt.
\label{eq:b8}
\end{equation}

Therefore, we obtain
\begin{equation}
 V(b) \propto b^{p+q-3}
\label{eq:b9}
\end{equation}
for
\begin{equation}
 \frac{3}{2} < p+q < 3.
\label{eq:b10}
\end{equation}
The visibilities of an intensity distribution with a spherically symmetric 
power law, $I\propto r_{\rm b}^{-X}$ for $1/2 < X < 2$, are a power law in 
the $u$--$v$ domain, $V \propto b^{X-2}$.

\clearpage

\begin{deluxetable}{cccccccccccccccc}
\tabletypesize{\footnotesize}
\rotate
\tablecaption{Summary of 45 m Telescope Observational Parameters}
\tablewidth{0pt}
\tablehead{
\colhead{Emission Line} & 
\colhead{$\nu$\tablenotemark{a}} & 
\colhead{Receiver} & 
\colhead{$\theta_{\rm HPBW}$\tablenotemark{b}} & 
\colhead{$\eta_{\rm mb}$\tablenotemark{c}} & 
\colhead{$\Delta v_{\rm res}$\tablenotemark{d}} &
\colhead{$\sigma_{T_A^\ast}$\tablenotemark{e}} & 
\colhead{Mode\tablenotemark{f}} & 
\colhead{Area\tablenotemark{g}} \\
\colhead{} & 
\colhead{(GHz)} & 
\colhead{} & 
\colhead{(\arcsec)} & 
\colhead{} & 
\colhead{(${\rm km\,s^{-1}}$)} & 
\colhead{(mK)} & 
\colhead{} & 
\colhead{(\arcmin)} & 
}
\startdata
${\rm NH_3}(J,K)=(1,1)$\tablenotemark{h} & 
$23.694506$ & 
H22 & 
$78$ & 
$0.82$ & 
$0.477$ & 
$31$ & 
PS & 
C \\
${\rm NH_3}(J,K)=(2,2)$\tablenotemark{h} & 
$23.722634$ & 
H22 & 
$78$ & 
$0.82$ & 
$0.476$ & 
$32$ & 
PS & 
C \\
${\rm NH_3}(J,K)=(3,3)$\tablenotemark{h} & 
$23.870130$ & 
H22 & 
$78$ & 
$0.82$ & 
$0.473$ & 
$35$ & 
PS & 
C \\
${\rm H^{13}CO^+}(J=1$--$0)$ & 
$86.75433$ & 
BEARS & 
$18.5$ & 
$0.5$ & 
$0.108$ &
$95$ & 
OTF & 
$4\times 4$\\ 
\enddata
\tablenotetext{a}{Rest frequency.}
\tablenotetext{b}{Half-power beam width for a Gaussian beam.}
\tablenotetext{c}{Main-beam efficiency.}
\tablenotetext{d}{Velocity resolution.}
\tablenotetext{e}{Typical rms noise level of the spectrum.}
\tablenotetext{f}{Observing mode; PS denotes the position-switching 
observations and OTF denotes the On-The-Fly observing mode.}
\tablenotetext{g}{Size of the region for the mapping observations. ``C''
denotes the one-point observation toward the IRAS source at the core 
center.}
\tablenotetext{h}{Emission lines of three transitions were obtained
simultaneously.}
\label{tbl:1}
\end{deluxetable}

\clearpage

\begin{deluxetable}{cccccccc}
\tabletypesize{\scriptsize}
\rotate
\tablecaption{Summary of NMA Observational Parameters}
\tablewidth{0pt}
\tablehead{
\colhead{Emission Line} & 
\colhead{$\nu$\tablenotemark{a}} & 
\colhead{Configuration} &
\colhead{Phase Reference Center} &
\colhead{$\theta_{\rm pri}$\tablenotemark{b}} &
\colhead{$\Delta v_{\rm res}$\tablenotemark{c}} &
\colhead{Gain Calibrator} &
\colhead{Passband Calibrator}\\ 
\colhead{} &
\colhead{(GHz)} & 
\colhead{($u$--$v$ Range (${\rm k\lambda}$))} &
\colhead{(B1950)} &
\colhead{(\arcsec)} &
\colhead{(${\rm km\,s^{-1}}$)} &
\colhead{} &
\colhead{}
}
\startdata
${\rm H^{13}CO^+}(J=1$--$0)$ & 
$86.75433$ & 
D and C ($2.89$--$47.2$) & 
19:34:35.1, 07:27:22.0 & 
78.9 & 
0.108 &
B1923+210 & 
3C345,\,3C454.3\\
\enddata
\tablenotetext{a}{Rest frequency.}
\tablenotetext{b}{Primary beam size which is defined as full width at half-maximum for a circular Gaussian pattern.}
\tablenotetext{c}{Velocity resolution.}
\label{tbl:2}
\end{deluxetable}

\clearpage

\begin{deluxetable}{ccccccc}
\tablewidth{0pt}
\tablecaption{Properties of ${\rm NH_3}$ Line Main Components}
\tabletypesize{\small}
\tablehead{
\multicolumn{3}{c}{${\rm NH_3}(1,1)$ \tablenotemark{a}} & & \multicolumn{3}{c}{${\rm NH_3}(2,2)$ \tablenotemark{a}}\\
\cline{1-3} \cline{5-7}\\
\colhead{$T_{\rm mb}$ \tablenotemark{b}} &
\colhead{$v_{\rm LSR}$ \tablenotemark{c}} &
\colhead{$\Delta v$ \tablenotemark{d}} &
&
\colhead{$T_{\rm mb}$ \tablenotemark{b}} &
\colhead{$v_{\rm LSR}$ \tablenotemark{c}} &
\colhead{$\Delta v$ \tablenotemark{d}}\\
\colhead{(K)} &
\colhead{(${\rm km\,s^{-1}}$)} &
\colhead{(${\rm km\,s^{-1}}$)} &
&
\colhead{(K)} &
\colhead{(${\rm km\,s^{-1}}$)} &
\colhead{(${\rm km\,s^{-1}}$)}
}
\startdata
$0.79\pm 0.06$ & $8.48\pm0.05$ & $0.99\pm 0.08$ & & $0.24\pm0.07$ & $8.36\pm 0.08$ & $0.76\pm0.13$\\
\enddata
\tablenotetext{a}{{}Line properties for the brightest hyperfine components are shown.}
\tablenotetext{b}{Peak main-beam brightness temperature.}
\tablenotetext{c}{{}LSR velocity at the peak brightness temperature by Gaussian fitting.}
\tablenotetext{d}{Velocity FWHM by Gaussian fitting.}
\label{tbl:3}
\end{deluxetable}

\clearpage

\clearpage

\clearpage

\clearpage

\clearpage

\clearpage

\clearpage

\clearpage

\begin{figure}[htbp]
\includegraphics[width=.9\linewidth]{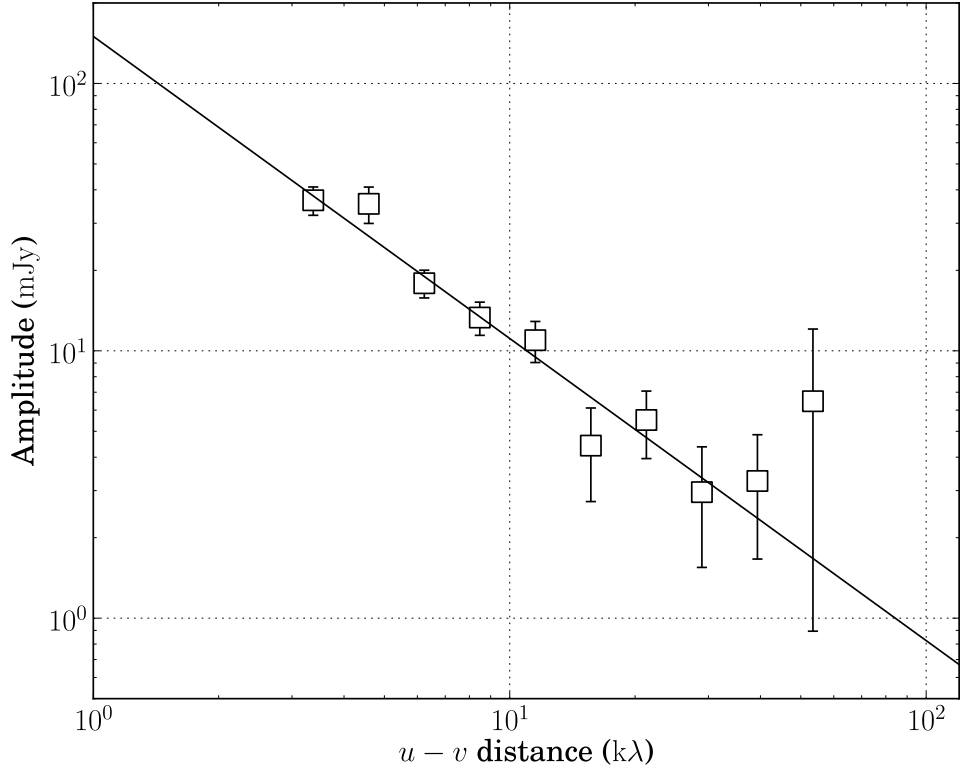}
\caption[
Binned visibility amplitude as a function of $u$-$v$ distance.
]
{
Binned visibility amplitude of the $87\,{\rm GHz}$ continuum data 
taken with the NMA as a function of $u$--$v$ distance.
Binning is logarithmic and amplitudes are obtained as vectorial average 
of the complex visibilities in each bin.
A solid line shows the best-fit power-law function.
}
\label{fig:8}
\end{figure}

\clearpage

\begin{figure}[htbp]
\includegraphics[width=.9\linewidth]{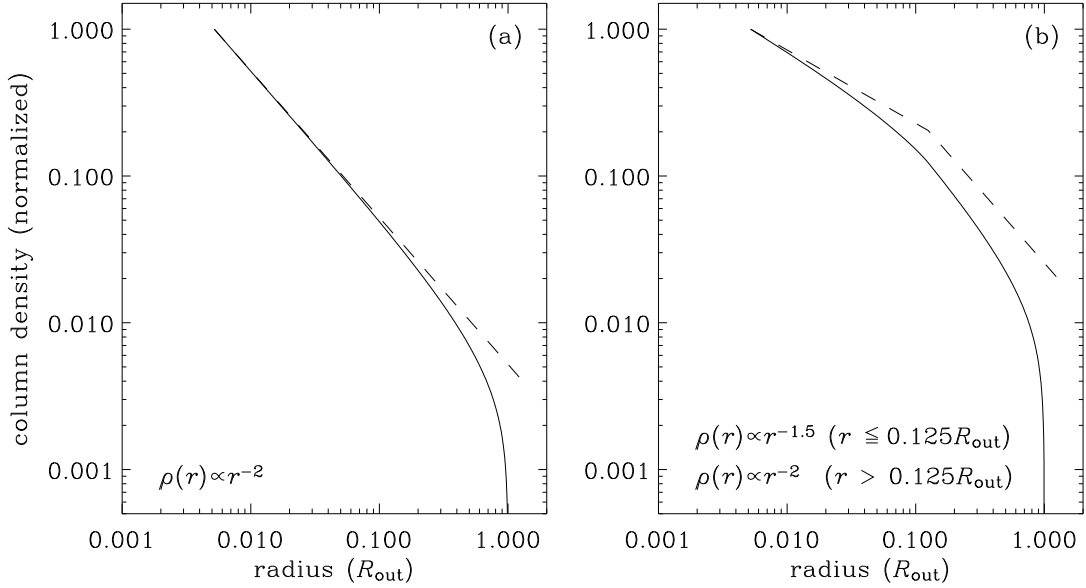}
\caption[
Column density profiles for power-law density distributions with a 
cutoff radius $R_{\rm out}$.
]
{
Column density profiles of core having power-law density distributions 
with a cutoff radius of $R_{\rm out}$.
Panel (a) shows the case of a density distribution of 
$\rho (r)\propto r^{-2}$ and
panel (b) shows the case of a density distribution of 
$\rho (r)\propto r^{-1.5}$ 
($r\leqslant 0.125 R_{\rm out}$) and $\rho (r)\propto r^{-2}$ 
($r> 0.125R_{\rm out}$).
Dashed lines indicate column density profiles expected from simple 
predictions of $\propto r^{-1}$ for (a), and $\propto r^{-0.5}$ 
($r\leqslant 0.125
R_{\rm out}$) and $\propto r^{-1}$ ($r> 0.125R_{\rm out}$) for (b).
}
\label{fig:9}
\end{figure}

\clearpage

\begin{figure}[htbp]
\includegraphics[width=\linewidth]{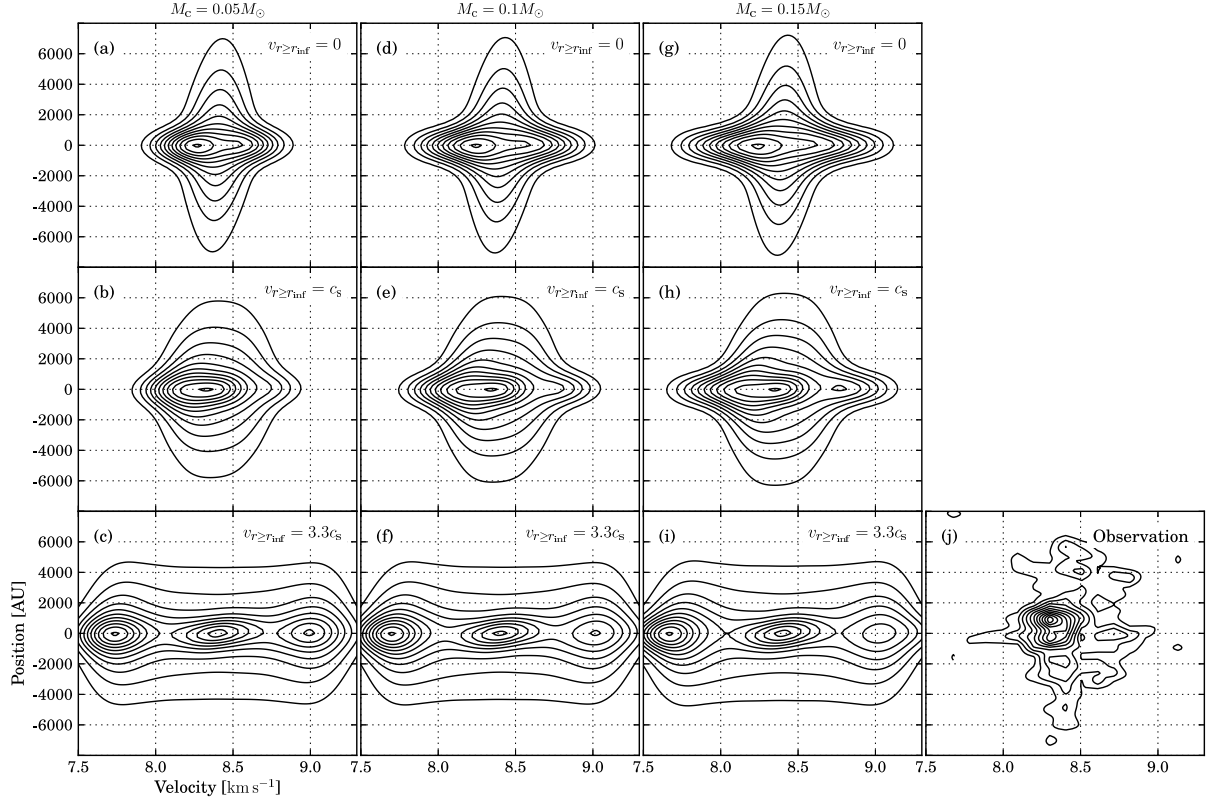}
\caption[
PV diagram of the ${\rm H^{13}CO^+}(J=1$--$0)$ line emission from B335
perpendicular to the outflow axis (black contours), overlaid by our 
simple calculations using models of contracting core with rotational 
motion.
]
{
Panels (a)--(i) show simulated PV diagrams by the simple 
calculations using models of contracting core with rotational motion.
The cutting axis is perpendicular to the rotation axis.
Panels (a), (b), and (c) show the model PV 
diagrams with 
$v_{r\geqslant r_{\rm inf}}= 0\,{\rm km\,s^{-1}}$, $c_{\rm s}$, and 
$3.3\,c_{\rm s}$ with the central stellar mass of $M_{\rm c}=0.05M_\sun$,
respectively. 
Panels (d)--(f) show the diagrams in the cases of
$M_{\rm c}=0.1M_\sun$ and panels (g)--(i) show the 
cases of $M_{\rm c}=0.15M_\sun$. 
PV diagram of the ${\rm H^{13}CO^+}(J=1$--$0)$ line emission from B335
perpendicular to the outflow axis is shown in panel (j) with
black contours.
Contour intervals of model PV diagrams relative to the peaks are the same 
as those for the observed diagram.
}
\label{fig:10}
\end{figure}

\clearpage

\begin{figure}[htbp]
\includegraphics[width=.7\linewidth]{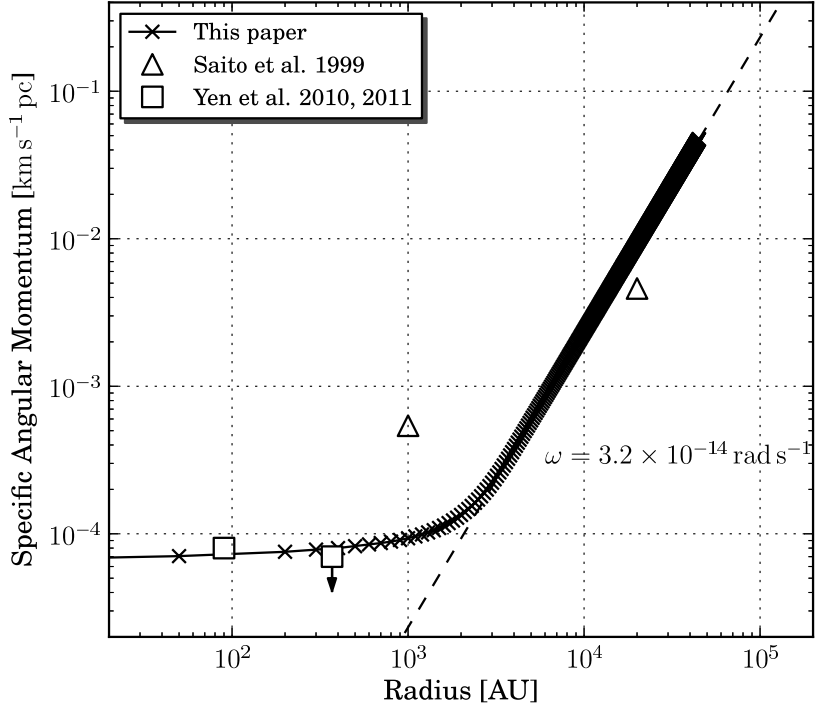}
\caption[
Radial distribution of specific angular momentum in our model 
calculation, measured by \citet{2010ApJ...710.1786Y,2011ApJ...742...57Y},
and by \citet{1999ApJ...518..334S}, as a function of radius.
]
{
Radial distribution of specific angular momentum in our model 
calculation (Section \ref{subsec:4.2}) for the case of $v_{r\geqslant r_{\rm inf}}= 0\,{\rm km\,s^{-1}}$ 
and $M_{\rm c}=0.1M_\sun$ (cross marks with solid line) as a function 
of radius.
Open squares represent specific angular momenta measured from velocity 
gradients seen in the ${\rm C^{18}O}(J=2$--$1)$ and ${\rm CS}(J=7$--$6)$ 
envelopes by \citet{2010ApJ...710.1786Y,2011ApJ...742...57Y}, and
open triangles denote specific angular momenta measured from 
${\rm C^{18}O}(J=1$--$0)$ core and 
${\rm H^{13}CO^+}(J=1$--$0)$ envelope by \citet{1999ApJ...518..334S}.
A dashed line indicates the profile of specific angular momentum of
$\omega = 3.2\times10^{-14}\,{\rm rad\,s^{-1}}$.
}
\label{fig:11}
\end{figure}

\end{document}